\documentclass[fleqn,12pt]{wlscirep}
\usepackage[utf8]{inputenc}
\usepackage[T1]{fontenc}
\usepackage[numbers,super]{natbib}
\usepackage{caption}

\newcommand{\araa}{Annu. Rev. Astron. Astrophys.}   
\newcommand{\aj}{Astron. J.}   
\newcommand{\apj}{Astrophys. J.}   
\newcommand{\apjl}{Astrophys. J. Lett.}   
\newcommand{\apjs}{Astrophys. J. Suppl. Ser.}   
\newcommand{\aap}{Astron. Astrophys.}   
\newcommand{\mnras}{Mon. Not. R. Astron. Soc.}   
\newcommand{\nat}{Nature} 
\newcommand{\pasj}{Publ. Astron. Soc. Jpn}   

\newcommand{\arcsec}{\mbox{\ensuremath{^{\prime\prime}}}}
\newcommand{\farcs}{\mbox{\ensuremath{.\!\!^{\prime\prime}}}}

\def\mbh{M$_{\rm BH}$}

\newcommand{\hst}{{\it HST}}
\newcommand{\jw}{{\it JWST}}
\newcommand{\targetone}{{J2255+0251}} 
\newcommand{\targettwo}{{J2236+0032}}
\newcommand{\galight}{\texttt{galight}}
\newcommand{\sersic}{S\'ersic}
\newcommand{\reff}{R$_{\rm eff}$}
\newcommand{\av}{A$_{\rm v}$}
\usepackage{color,subfigure}

\graphicspath{{./}{figures/}}

\title{Detection of stellar light from quasar host galaxies at redshifts above 6}

\author[1,2,*]{Xuheng Ding}
\author[3,1,4,*]{Masafusa Onoue}
\author[1,2,5]{John D. Silverman}
\author[6]{Yoshiki Matsuoka}
\author[7,8]{Takuma Izumi}
\author[9]{Michael A. Strauss}
\author[4]{Knud Jahnke}
\author[9]{Camryn L. Phillips}
\author[10]{Junyao Li}
\author[11]{Marta Volonteri}
\author[12,13]{Zoltan Haiman}
\author[14,15]{Irham Taufik Andika}
\author[16]{Kentaro Aoki}
\author[17]{Shunsuke Baba}
\author[18]{Rebekka Bieri}
\author[4]{Sarah E. I. Bosman}
\author[1,2]{Connor Bottrell}
\author[19]{Anna-Christina Eilers}
\author[20]{Seiji Fujimoto}
\author[21,4]{Melanie Habouzit}
\author[7,22]{Masatoshi Imanishi}
\author[3]{Kohei Inayoshi}
\author[23,24]{Kazushi Iwasawa}
\author[5,25]{Nobunari Kashikawa}
\author[26]{Toshihiro Kawaguchi}
\author[27,25]{Kotaro Kohno}
\author[28]{Chien-Hsiu Lee}
\author[29]{Alessandro Lupi}
\author[30]{Jianwei Lyu}
\author[6]{Tohru Nagao}
\author[31]{Roderik Overzier}
\author[32]{Jan-Torge Schindler}
\author[33]{Malte Schramm}
\author[5,25]{Kazuhiro Shimasaku}
\author[7,34]{Yoshiki Toba}
\author[35]{Benny Trakhtenbrot}
\author[36]{Maxime Trebitsch}
\author[37]{Tommaso Treu}
\author[38,39]{Hideki Umehata}
\author[32]{Bram P. Venemans}
\author[30,40]{Marianne Vestergaard}
\author[4]{Fabian Walter}
\author[30]{Feige Wang}
\author[30]{Jinyi Yang}

\affil[1]{Kavli Institute for the Physics and Mathematics of the Universe (Kavli IPMU, WPI), The University of Tokyo, Chiba 277-8583, Japan }
\affil[2]{Center for Data-Driven Discovery, Kavli IPMU (WPI), UTIAS, The University of Tokyo, Kashiwa, Chiba 277-8583, Japan}
\affil[3]{Kavli Institute for Astronomy and Astrophysics, Peking University, Beijing 100871, China }
\affil[4]{Max-Planck-Institut f\"{u}r Astronomie, K\"{o}nigstuhl 17, D-69117 Heidelberg, Germany }
\affil[5]{Department of Astronomy, School of Science, The University of Tokyo, 7-3-1 Hongo, Bunkyo-ku, Tokyo, 113-0033, Japan }
\affil[6]{Research Center for Space and Cosmic Evolution, Ehime University, 2-5 Bunkyo-cho, Matsuyama, Ehime 790-8577, Japan }
\affil[7]{National Astronomical Observatory of Japan, Osawa, Mitaka, Tokyo 181-8588, Japan }
\affil[8]{Department of Physics, Graduate School of Science, Tokyo Metropolitan University, 1-1 Minami-Osawa, Hachioji, Tokyo 192-0397, Japan }
\affil[9]{Department of Astrophysical Sciences, Princeton University, 4 Ivy Lane, Princeton, NJ 08544, USA}
\affil[10]{Department of Astronomy, University of Illinois at Urbana-Champaign, Urbana, IL, 61801, USA}
\affil[11]{Institut d'Astrophysique de Paris, CNRS, Sorbonne Universit\'e, UMR7095, 98bis bd Arago, 75014 Paris, France }
\affil[12]{Department of Astronomy, Columbia University,  New York, NY 10027, USA }
\affil[13]{Department of Physics, Columbia University,  New York, NY 10027, USA }
\affil[14]{Physik-Department, Technische Universit\"{a}t M\"{u}nchen, James-Franck-Str.~1, D-85748 Garching bei M\"{u}nchen, Germany }
\affil[15]{Max-Planck-Institut f\"{u}r Astrophysik, Karl-Schwarzschild-Str.~1, D-85748 Garching bei M\"{u}nchen, Germany }
\affil[16]{Subaru Telescope, National Astronomical Observatory of Japan,  650 North A'ohoku Place, Hilo , Hawaii 96720 U.S.A. }
\affil[17]{Graduate School of Science and Engineering, Kagoshima University, 1-21-35 Korimoto, Kagoshima, Kagoshima 890-0065, Japan }
\affil[18]{Institute for Computational Science, University of Zurich, Wintherthurerstrasse 190, 8057 Z\"urich, Switzerland}
\affil[19]{MIT Kavli Institute for Astrophysics and Space Research, 77 Massachusetts Avenue, Cambridge, 02139, Massachusetts, USA }
\affil[20]{Department of Astronomy, The University of Texas at Austin, Austin, TX 78712, USA }
\affil[21]{Zentrum f\"ur Astronomie der Universit\"at Heidelberg (ITA), Albert-Ueberle-Str. 2, D-69120 Heidelberg, Germany }
\affil[22]{Department of Astronomy, School of Science, Graduate University for Advanced Studies (SOKENDAI), Mitaka, Tokyo 181-8588, Japan }
\affil[23]{Institut de Ci\`encies del Cosmos (ICCUB), Universitat de Barcelona (IEEC-UB), Mart\'i i Franqu\`es, 1, 08028 Barcelona, Spain}
\affil[24]{ICREA, Pg. Llu\'is Companys 23, 08010 Barcelona, Spain}
\affil[25]{Research Center for the Early Universe, Graduate School of Science, The University of Tokyo, 7-3-1 Hongo, Bunkyo-ku, Tokyo 113-0033, Japan }
\affil[26]{Department of Economics, Management and Information Science, Onomichi City University, Hisayamada 1600-2, Onomichi, Hiroshima 722-8506, Japan }
\affil[27]{Institute of Astronomy, Graduate School of Science, The University of Tokyo, 2-21-1 Osawa, Mitaka, Tokyo 181-0015, Japan }
\affil[28]{W. M. Keck Observatory, 65-1120 Mamalahoa Hwy, Kamuela, HI 96743, USA}
\affil[29]{Dipartimento di Fisica ``G. Occhialini'', Universit\`a degli Studi di Milano-Bicocca, Piazza della Scienza 3, I-20126 Milano, Italy }
\affil[30]{Steward Observatory, University of Arizona, 933 N. Cherry Avenue, Tucson AZ 85721, USA }
\affil[31]{Observat\'orio Nacional/MCTI, Rua General Jos\'e Cristino, 77, S\~{a}o Crist\'ov\~{a}o, Rio de Janeiro, RJ 20921-400, Brazil }
\affil[32]{Leiden Observatory, Leiden University, PO Box 9513, 2300 RA Leiden, The Netherlands }
\affil[33]{Universit\"{a}t Potsdam, Karl-Liebknecht-Str. 24/25, D-14476 Potsdam, Germany }
\affil[34]{Academia Sinica Institute of Astronomy and Astrophysics, 11F Astronomy-Mathematics Building, AS/NTU, No.1, Section 4, Roosevelt Road, Taipei 10617, Taiwan }
\affil[35]{School of Physics and Astronomy, Tel Aviv University, Tel Aviv 69978, Israel }
\affil[36]{Kapteyn Astronomical Institute, University of Groningen, P.O. Box 800, 9700 AV Groningen, The Netherlands }
\affil[37]{Department of Physics and Astronomy, University of California Los Angeles, CA, 90095, USA }
\affil[38]{Institute for Advanced Research, Nagoya University, Furocho, Chikusa, Nagoya 464-8602, Japan }
\affil[39]{Department of Physics, Graduate School of Science, Nagoya University, Furocho, Chikusa, Nagoya 464-8602, Japan }
\affil[40]{DARK, Niels Bohr Institute, Jagtvej 155, 2200 Copenhagen N, Denmark}

\affil[*]{Corresponding author}

\makeatletter
\renewcommand{\@maketitle}{%
{%
\thispagestyle{empty}%
\vskip-36pt%
{\raggedright\sffamily\bfseries\fontsize{20}{25}\selectfont \@title\par}%
\vskip10pt
{\raggedright\sffamily\fontsize{12}{16}\selectfont  \@author\par}
\vskip25pt%
}%
}%
\makeatother

\begin{document}

\flushbottom
\maketitle

\noindent\textbf{The detection of starlight from the host galaxies of quasars during the reionization epoch ($z>6$) has been elusive, even with deep \hst\ observations\cite{Mechtley2012,Marshall2020}. The current highest redshift quasar host detected\cite{Peng2006}, at $z=4.5$, required the magnifying effect of a foreground lensing galaxy.  Low-luminosity quasars\cite{Matsuoka2016, Matsuoka2018, Matsuoka2022} from the Hyper Suprime-Cam Subaru Strategic Program (HSC-SSP)\cite{HSC-SSP2018} mitigate the challenge of detecting their underlying, previously-undetected host galaxies. Here we report rest-frame optical images and spectroscopy of two HSC-SSP quasars at $\mathbf{z>6}$ with \jw. Using NIRCam imaging at 3.6$\mu$m and 1.5$\mu$m and subtracting the light from the unresolved quasars, we find that the host galaxies are massive (stellar masses of $\mathbf{13\times}$ and $\mathbf{3.4}\times$ $\mathbf{10^{10}}$ M$_{\odot}$, respectively), compact, and disk-like. NIRSpec medium-resolution spectroscopy shows stellar absorption lines in the more massive quasar, confirming the detection of the host.  Velocity-broadened gas in the vicinity of these quasars enables measurements of their black hole masses ($\mathbf{1.4\times 10^9}$ and $\mathbf{2.0}\times$ $\mathbf{10^{8}}$ M$_{\odot}$, respectively). Their location in the black hole mass - stellar mass plane is consistent with the distribution at low redshift, suggesting that the relation between black holes and their host galaxies was already in place less than a billion years after the Big Bang.
} \\

The quasars \targettwo\ and \targetone\ (coordinates listed in Extended Data Table~\ref{tab:table1}) are at redshifts of $6.40$ and $6.34$.  They have relatively low luminosity ($M_{1450}=-23.87$ and $-23.66$ mag, respectively), roughly 10 times less luminous than typical bright quasars at these redshifts\cite{Fan2001, 2018Natur.553..473B, Fan2022}.
Using \jw\ with NIRCam\cite{Rieke2005}, we observed \targettwo\ on November 6, 2022 and  \targetone\ on October 26, 2022. Images were taken in two broad-band filters (F356W and F150W) to detect their host galaxies at 3.56 and 1.50 $\mu$m, thus bracketing the rest-frame \hbox{4000 \AA} break, a characteristic feature of galaxy spectra which constrains the age of the stellar population. Total exposure times were approximately 3,100 seconds for each filter per target.

\begin{figure}[ht]
\vspace{-0.5cm}
\centering
\includegraphics[trim = 0mm 20mm 0mm 15mm, clip, width=0.9\textwidth]{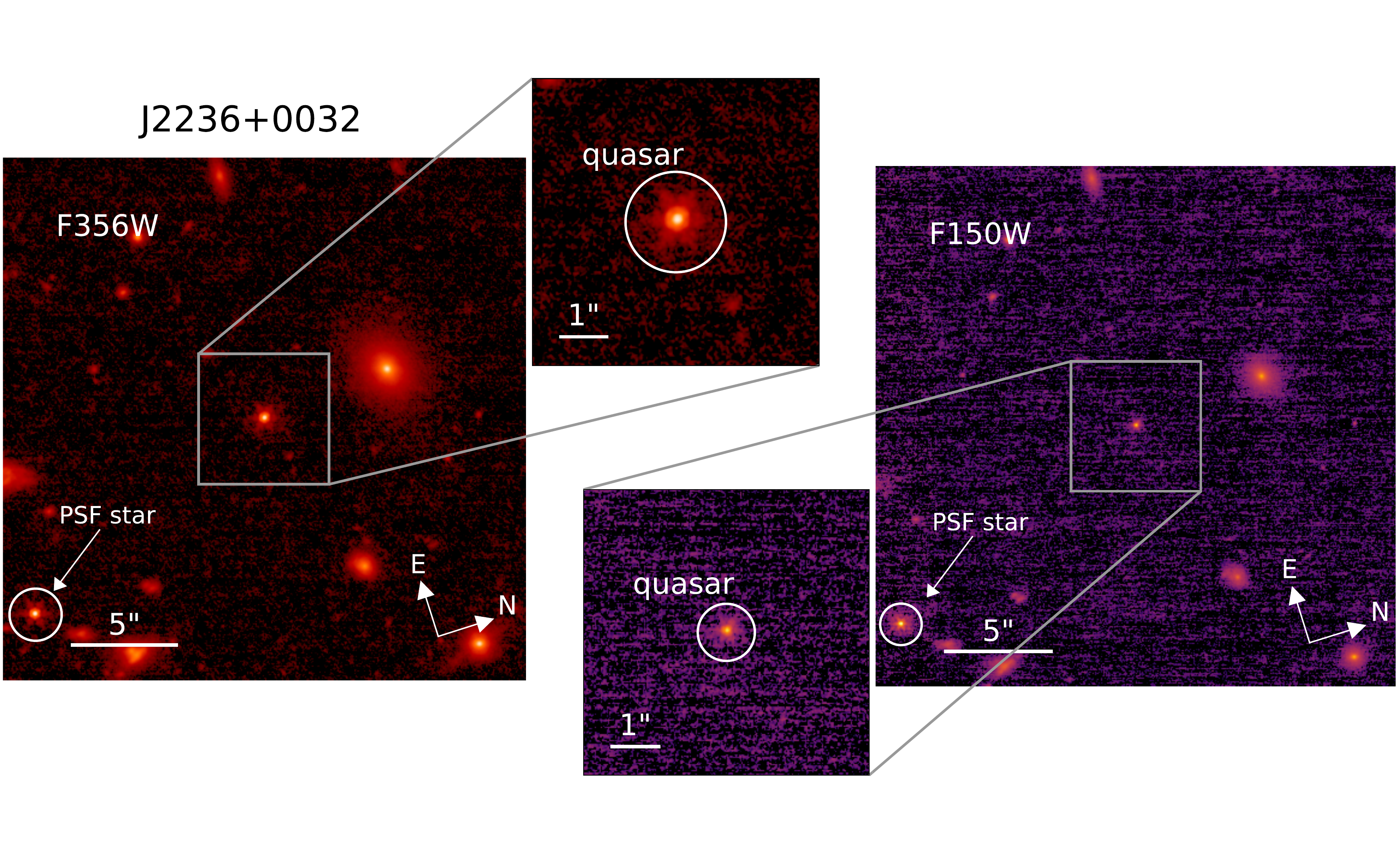}\\
\includegraphics[trim = 0mm 20mm 0mm 15mm, clip, width=0.9\textwidth]{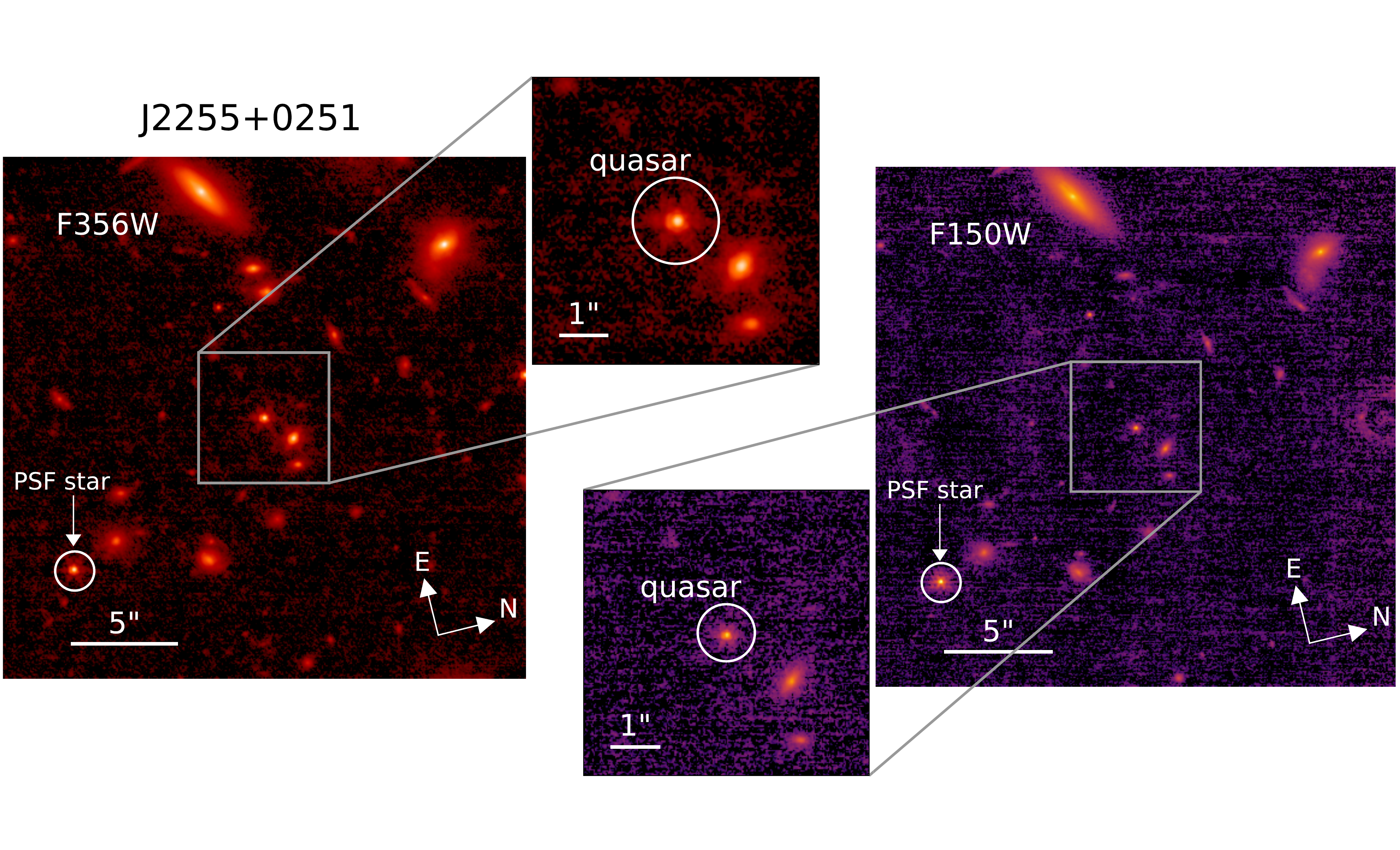}
\caption{\textbf{NIRCam images of the field around \targettwo\ and \targetone.} Both JWST filters F356W ({\it left panels}) and F150W ({\it right panels}) are shown. 
The middle panels are zoomed-in regions of each image. One of the PSF stars used for 2D image decomposition is indicated in each quasar field. 
} \label{fig:fov}
\end{figure}

In Figure~\ref{fig:fov}, we show a $24$\arcsec$\times24$\arcsec\ region and zoomed-in cutouts of the \jw\ images centered on the quasars for both filters. As expected, both quasars are detected at high significance. The first quasar \targettwo\ ($m_{\rm F356W}=21.75 \pm 0.02$ mag, $m_{\rm F150W}=22.78 \pm 0.02$,) has no evidence of close neighbors. The second quasar \targetone\ ($m_{\rm F356W}=22.20 \pm 0.02$ mag, $m_{\rm F150W}=23.03 \pm 0.02$) has two neighboring sources ($<2.6^{\prime\prime}$) that do not have spectroscopic redshifts at this point; however, their detection in the HSC $r$- and $i$-bands suggests that both of them are at lower redshift than the central quasar.

The detection of an underlying host galaxy requires careful decomposition of the two-dimensional light distribution to separate the unresolved quasar from its host galaxy. We perform a two-component model fit of the rest-frame optical emission in a region of the NIRCam images centered on each target. We use the software \galight\cite{Ding2020} to identify stars from each image to model the two-dimensional (2D) point-spread function (PSF).  We then model each quasar image in each filter as a sum of a point-like quasar and an extended host. 
The host galaxy was modeled using a 2D \sersic\ profile\cite{Sersic1963} whose parameters include the position, size (\reff), axis ratio, orientation, flux, and the radial profile/shape of the light distribution (i.e., \sersic\ index), allowing a flexible description of the {\em a priori} unknown host galaxy shape. In Extended Data Table~\ref{tab:table1}, we report the best-fit parameters and their uncertainties based on the dispersion resulting from different PSF models with different modeling techniques (see Methods).

\begin{figure}
\centering
\vspace{-1.5cm}
\includegraphics[trim = 70mm 5mm 73mm 1mm, clip, width=0.85\textwidth]{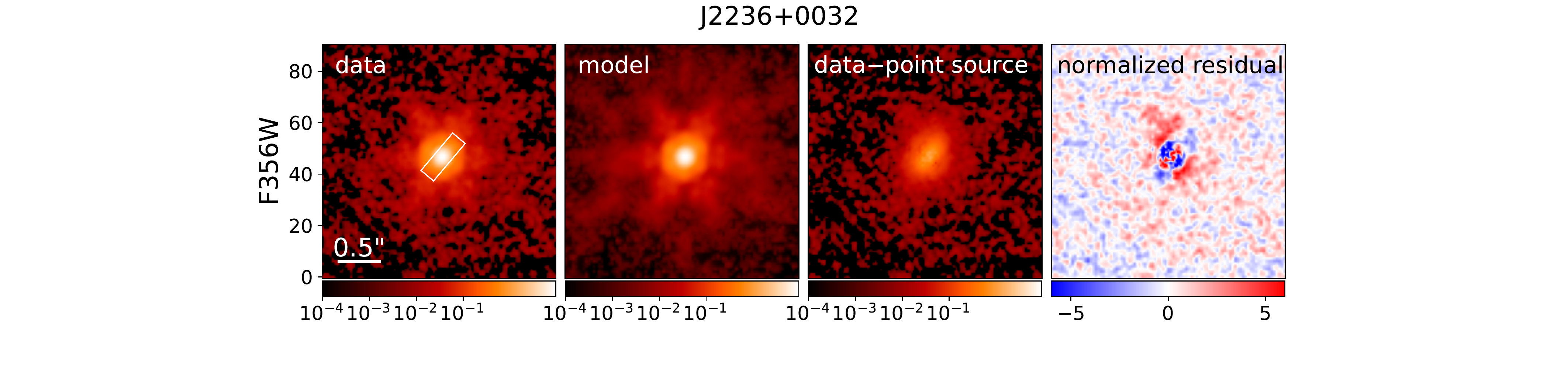}\\
\vspace{-0.3cm}
\includegraphics[trim = 70mm 5mm 73mm 10mm, clip, width=0.85\textwidth]{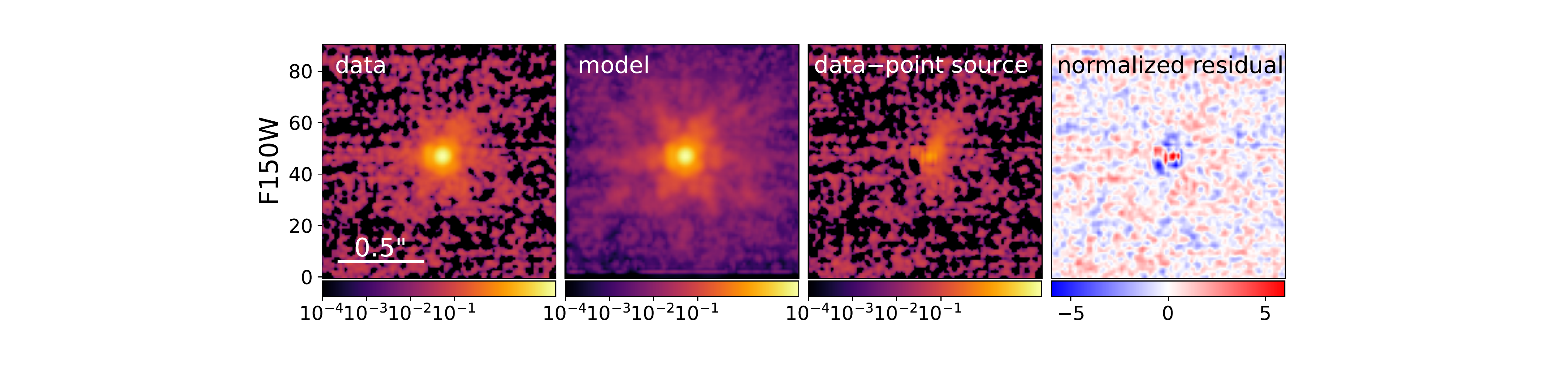}\\

\includegraphics[trim = 70mm 5mm 73mm 1mm, clip, width=0.85\textwidth]{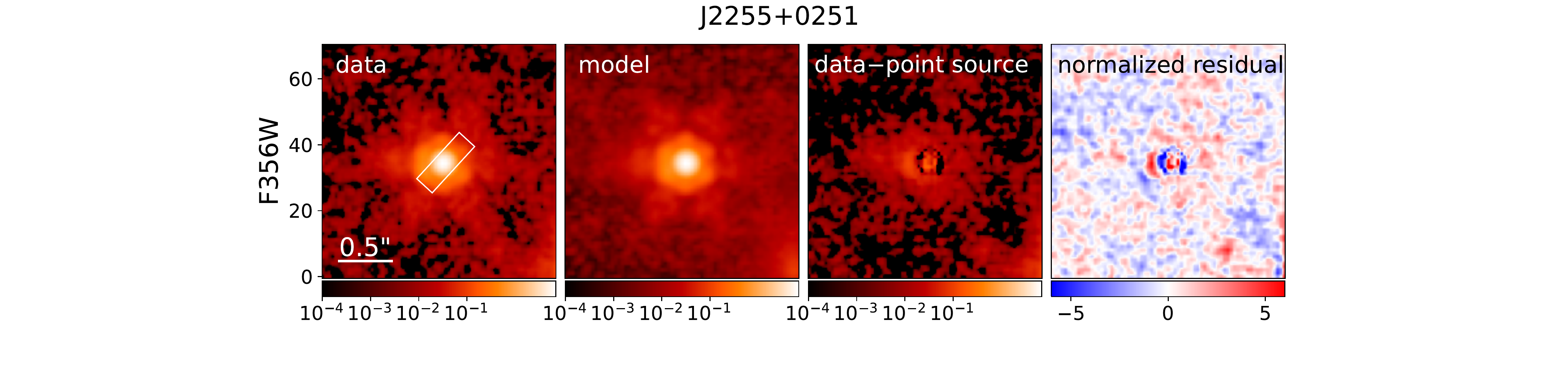}\\
\vspace{-0.3cm}
\includegraphics[trim = 70mm 5mm 73mm 10mm, clip, width=0.85\textwidth]{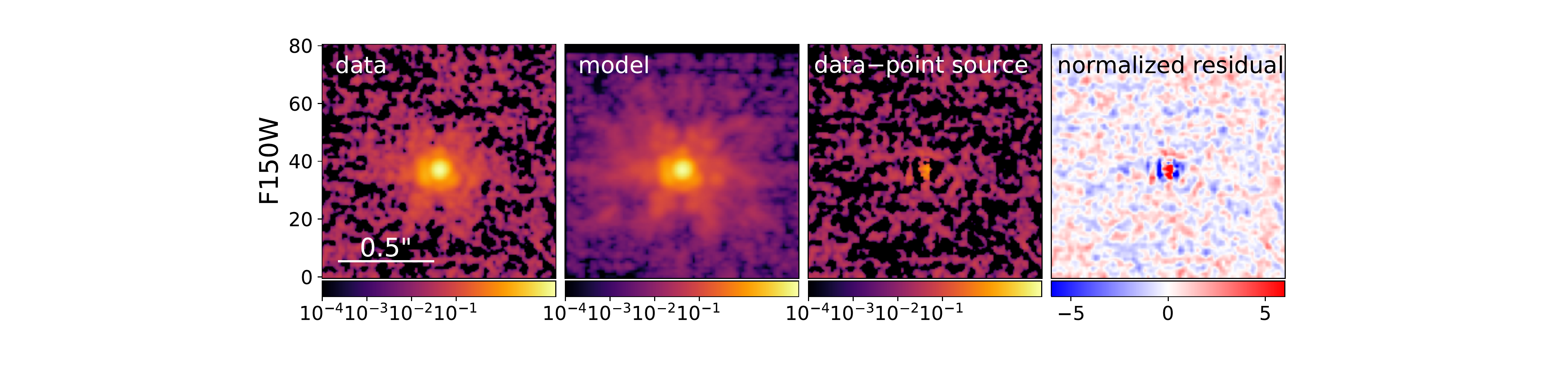}\\

\caption{\textbf{2D decomposition into quasar and host-galaxy emission.} Panels are as follows from left to right: quasar image (data), best-fit model (quasar + host galaxy), \underline{host galaxy only} (data $-$ model quasar point-source component), and normalized residual image, i.e., (data $-$ model) / $\sigma$, where $\sigma$ is the flux uncertainty of each pixel.
The target name is above each row of panels and the filter is indicated on the left. The alignment of the NIRSpec slit ($0\farcs{}2\times0\farcs{}6$) is shown in the left panel of the F356W image for each target. Pixel  units are Mega-Janskys per steradian. 
A scale bar of $0\farcs{}5$ corresponds to $2.8$~kpc at $z\sim6.4$.} \label{fig:decomp}
\end{figure}

As shown in Figure~\ref{fig:decomp}, we detect the host galaxies for both targets, which are visible after subtracting the quasar point-source model (third column). For \targettwo, the quasar host is significantly detected in both filters ($m_{\rm F356W}=23.12 \pm 0.20$ mag; 
$m_{\rm F150W}=25.12 \pm 0.29$ mag). The inferred host-to-total flux ratios are $25.5\pm4.4\%$, and $10.2\pm2.8\%$ with sizes (\reff) of $0.7\pm0.1$~kpc and $0.5\pm0.2$~kpc for F356W and F150W, respectively. The light distribution is elongated in the same direction (i.e., NE--SW) in the two filters, suggesting a mildly inclined disk-like host. For this target, we fix the \sersic\ index to $1$ in both bands -- corresponding to an exponential disk model -- to avoid non-physical fits; this is supported by a subsequent fit to the host galaxy emission after subtracting the quasar (see Methods).
For \targetone, the host galaxy is clearly detected in the F356W band ($m_{\rm F356W}= 24.58 \pm 0.30$ mag) with a host-to-total flux ratio of $9.8\pm2.6\%$. The host emission extends southeast (left) of the quasar's location. The size of the host is inferred to be \reff$=1.5\pm1.1$~kpc with a \sersic\ index of $1.5\pm1.4$, suggesting a more disk-like than spheroidal profile. The host galaxy is not detected in the F150W image, resulting in an upper limit of $m_{F150W}= 26.3$ mag (see Methods). This upper limit provides an important constraint on the galaxy spectral energy distribution (SED). A non-detection is not surprising, given the low host-to-total flux ratio in the F356W band and the fact that galaxies are typically fainter blueward of the 4000~\AA\ break. We use NIRSpec\cite{NIRSpec} observations to demonstrate that the contribution of extended line emission to the host photometry in F356W is minimal (See Methods). 

Our two photometric bands lie on either side of the 4000 \AA\ break, allowing us to quantify the contributions of young and relatively old stars, and thus to estimate the stellar masses of the two quasar host galaxies. 
We fit the SED of each host with a single stellar population,
in which the stellar metallicity $\log Z/Z_\odot$ is constrained to the range $[-1, -0.3]$, the stellar age is within the range $[0.01, 0.84]$~Gyr (capped by the age of the universe at redshift 6.4), and the extinction \av\ is in the range of $[0, 5.0]$ mag.
With standard assumptions on the stellar Initial Mass Function (IMF), we find that \targettwo\ has a stellar mass of 
$1.3\substack{+2.0\\-0.6}\times{10^{11}}\ {\rm M}_{\odot}$, while \targetone\ is less massive with $3.4\substack{+7.6\\-1.9}\times{10^{10}}\ {\rm M}_{\odot}$. We use simulations and three different SED fitting codes to demonstrate that the results are robust, although  further observations will lessen potential systematic errors in these mass estimates (see Methods for full details on our mass estimates and their uncertainties). Our host galaxies are among the highest stellar mass galaxies known at $z\sim6$ (ref. \citenum{Grazian2015}).

\begin{figure}
\centering
\includegraphics[trim = 10mm 40mm 8mm 40mm, clip, width=0.9\textwidth]{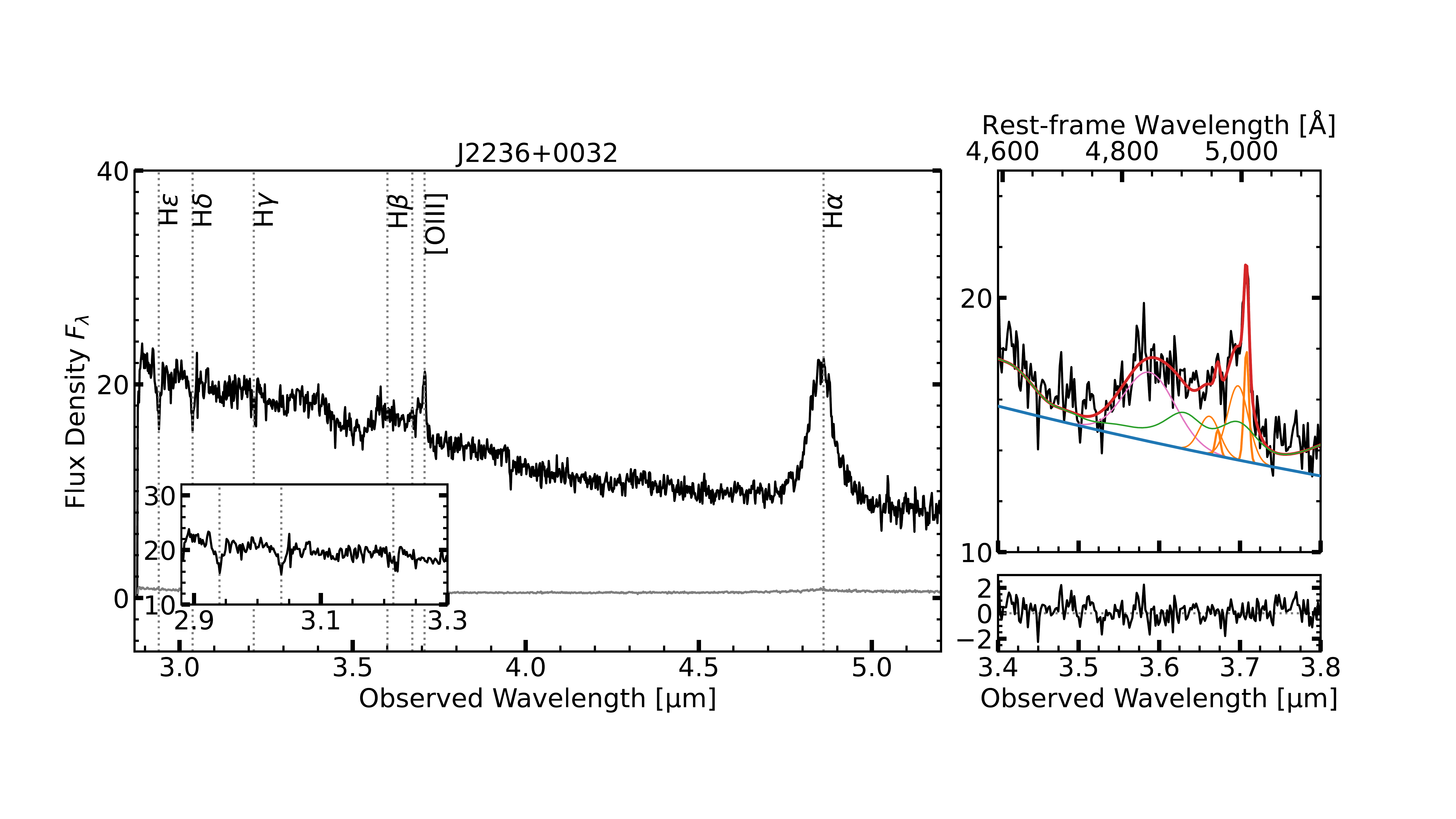}\\
\includegraphics[trim = 10mm 40mm 8mm 40mm, clip, width=0.9\textwidth]{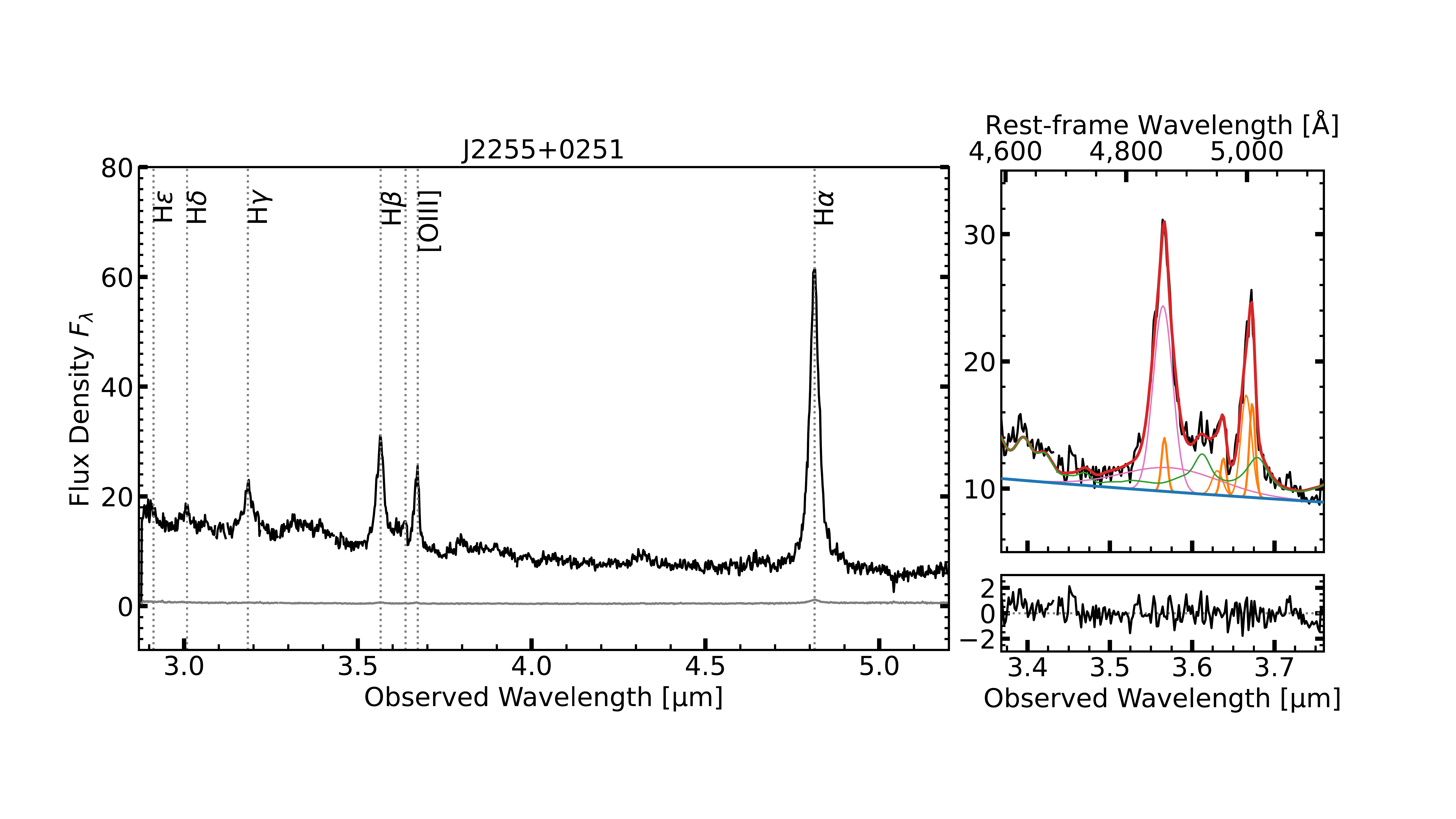}\\
\caption{ \textbf{NIRSpec rest-frame optical spectra of \targettwo\  and \targetone.}
For each quasar, the left panel shows the full spectrum while the top-right panels zoom in on the H$\beta$+[O~{\sc iii}] $\lambda\lambda\ 4960,5008$ doublet.
The flux density $F_\lambda$ is in units of $10^{-20}$ erg cm$^{-2}$ s$^{-1}$ \AA$^{-1}$.
The data are shown in black with the errors per pixel in grey. 
The stellar absorption lines in \targettwo\ are identified in H$\gamma$, H$\delta$ and H$\epsilon$, which are also presented in the inset zoom-in panel.
Best-fit models of power-law continuum (blue), Fe~{\sc ii} (green), broad H$\beta$ (magenta), narrow H$\beta$ + [O~{\sc iii}] (orange thick line), and broad [O~{\sc iii}] (orange thin line) are overplotted in the top-right panels.
The sum of the continuum and emission line models is shown in red.
The residuals are displayed in the bottom right panels.
}  \label{fig:NIRSpecData}
\end{figure}

A rest-frame optical spectrum of each quasar was obtained by the NIRSpec instrument on JWST (Figure~\ref{fig:NIRSpecData}).
These spectra cover 2.87--5.27 $\mu$m, or approximately 4000--7000 \AA\ in the rest-frame, revealing  strong emission lines including the H$\beta$+[O~{\sc iii}] multiplet, H$\alpha$, and blended Fe~{\sc ii} pseudo-continuum.
The 0\farcs2-wide slit was fortuitously aligned along the major axis of the host galaxy of \targettwo\ (Figure~\ref{fig:decomp}).
Remarkably, we serendipitously identify stellar absorption lines 
(H$\gamma$, H$\delta$, and H$\epsilon$) in \targettwo.  
Thus the contribution of stellar light is confirmed both from the imaging and spectroscopy in this object.
We use the velocity width of the broad H$\beta$ emission line and the rest-frame 5100 \AA\ continuum luminosity to estimate the mass of the central black hole,
under the assumption that the high-velocity gas is governed by the gravity of the SMBH.
Using a recipe calibrated in the local universe\cite{Vestergaard2006} and subtracting the stellar emission, we obtain virial masses of M$_{\rm BH} = 1.4 \times 10^9\ M_\odot$ for \targettwo, and M$_{\rm BH} = 2.0 \times 10^8\ M_\odot$ for \targetone, respectively (see Methods).

\begin{figure}
\centering
\hspace{-1.7cm}
\includegraphics[trim = 0mm 0mm 0mm 0mm, clip, width=\linewidth]{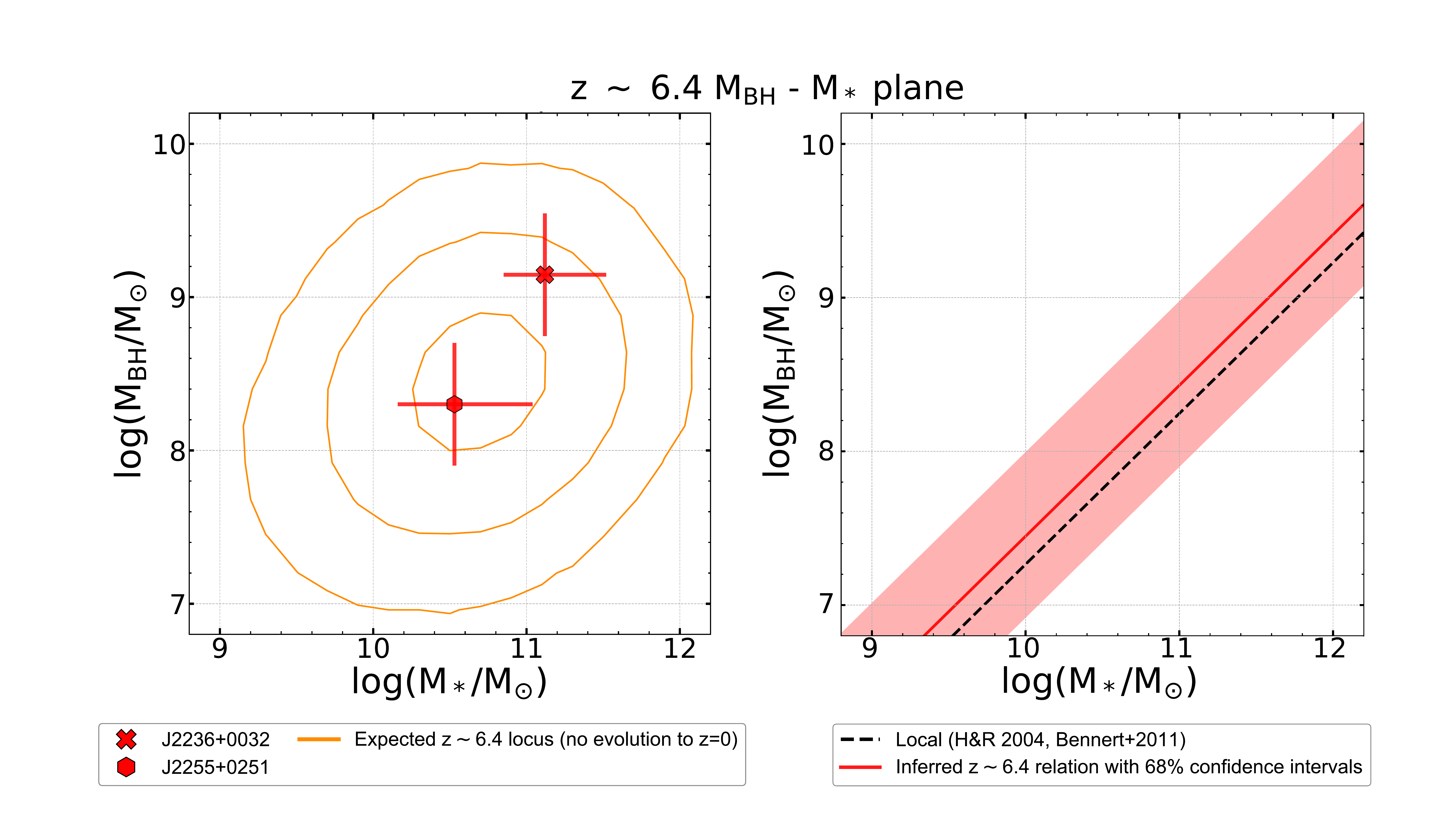} %
\caption{\textbf{BH mass -- stellar mass plane at $\mathbf{z\sim6.4}$.}
{\it Left:}  
Observed locations of our two $z\sim6.4$ quasars marked in red. Orange contours (1--3$\sigma$ levels) show the expected distribution of a mock quasar sample at $z\sim6.4$ for which the ratio M$_{\rm BH}$/M$_*$ of the intrinsic population does not evolve to $z=0$.
{\it Right:}
The M$_{\rm BH}$--M$_*$ relation at $z\sim6.4$ (red solid line) inferred from our two quasars after accounting for selection effects.  The shaded region represents the 68 \% confidence interval, and the dashed line is the observed relation at $z=0$\cite{HR2004, Bennert2011}.  Note that we assume a linear relationship between black hole mass and stellar mass in log-log space. The slope is fixed to the value observed at $z=0$.
}
\label{fig:Mbh_Mstellar}
\end{figure}

Figure~\ref{fig:Mbh_Mstellar} shows the two $z\sim6.4$ quasars in the black hole mass -- stellar mass plane (left panel), as well as the inferred underlying relation compared to the local relation (right panel). 
As is observed in the local universe, the quasar hosting the more massive black hole is hosted by the more massive galaxy. To evaluate their location in the M$_{\rm BH}$ -- M$_{\rm *}$ plane relative to the local mass relation, we carry out a simulation, following ref. \citenum{Li2022}, that takes into account the selection effects for these two quasars in the HSC survey. For this purpose, we combine the quasar flux limits, the estimated BH mass and accretion rate functions at $z\sim6$, as well as uncertainty in our stellar mass estimates to predict the expected M$_{\rm BH}$ for these two quasars -- under the baseline assumption that the M$_{\rm BH}$--M$_*$ relation does not evolve between $z=6.4$ and the present.   
In the left panel, the orange contours show the resulting bivariate probability distribution for the location of these two quasars, with the measured values overplotted in red.
Our two $z>6$ quasars fall within the probability distribution (orange contours) of this simulated $z\sim6.4$ sample and are thus consistent with no evolution of the M$_{\rm BH}$--M$_*$ relation from $z \sim 6.4$ to $z=0$, as shown for the inferred underlying relation in the right panel of Figure~\ref{fig:Mbh_Mstellar}. 
With just two quasars, the probability distribution of the nominal evolution strength is broad as suggested by the width of these uncertainties (see also Extended Data Figure~\ref{fig:gamma_prob} in the Methods). However, ongoing  \jw\ observations will give us a significantly larger sample, allowing us to better constrain models for the mutual evolution of the black hole and stellar populations in galaxies\cite{2013ARA&A..51..511K, Habouzit2022, Inayoshi2022}.

\newpage

\section*{Figures}

\newpage

\section*{Methods}

{\bf Cosmological model} 

\noindent A standard concordance cosmology with $H_0= 70$ km s$^{-1}$ Mpc$^{-1}$, $\Omega{_m} = 0.30$, and $\Omega{_\Lambda} = 0.70$ is adopted, which gives a scale of $5.63$ kpc/$\arcsec$ at $z=6.4$. All magnitudes are presented in the AB system. Milky Way dust extinction is negligible at the near-infrared wavelengths of interest and is not corrected for.  \\

\noindent {\bf Target selection} 

\noindent The two objects presented here are part of our Cycle 1 \jw\ program  (Observation ID 1967; PI: M.Onoue) which targets 12 of the lowest-luminosity quasars at redshift 6.0 to 6.4 (ref. \citenum{Matsuoka2018c}) discovered by the Hyper Suprime-Cam Subaru Strategic Program\cite{HSC-SSP2018}, an optical wide-field survey using the 8.2-meter Subaru telescope.
These quasars were spectroscopically confirmed by ground-based follow-up observations\cite{Matsuoka2016, Matsuoka2018, Matsuoka2022}.  The quasars exhibit broad Ly$\alpha$ emission and strong intergalactic medium absorption blueward of Ly$\alpha$. This deep survey, which is sensitive to quasars a factor of 10 less luminous than those discovered by shallower surveys such as the Sloan Digital Sky Survey (see Extended Data Figure~\ref{fig:shellqs_sample}). \\

\noindent {\bf NIRCam Observations and data reduction} 

\noindent The data presented in this paper were taken with Module B of the NIRCam instrument, which has a field of view of $2.2\times2.2$ square arcminutes.
Total exposures of 3,100 seconds in the two filters (F356W and F150W) were obtained simultaneously.
A $4\times4$ primary and sub-pixel dithering pattern was employed to mitigate cosmic ray hits and bad pixels in the detector and to ensure sub-pixel resampling during the stacking step.
We used the \textsf{INTRAMODULEBOX} and \textsf{STANDARD} dithering patterns for the primary and sub-pixel dithers, respectively.
We used the \textsf{BRIGHT1} readout mode.

The data were processed using the standard procedures in  \jw\ pipeline version 1.7.2.
The pre-calibrated ``Stage 2'' image frames were downloaded from the MAST archive.
These images have the pipeline parameter reference files \textsf{jwst\_1009.pmap} for \targetone\ and \textsf{jwst\_1011.pmap} for \targettwo, as registered in the \jw\ Calibration Reference Data System\footnote{\url{https://jwst-crds.stsci.edu}}.
For individual frames, global background light was first subtracted using the \textsf{Background2D} function of \textsf{Photutils}\cite{Bradley2022}.
The archived images clearly have horizontal and vertical stripe noise patterns, known as ``$1/f$ noise''.
This $1/f$ noise was subtracted by first masking bright objects, then collapsing the 2D images along each axis of the detectors and estimating the noise amplitudes by measuring sigma-clipped median values.  These amplitudes were then subtracted from each row and column. 
The horizontal stripes were measured for each of the four detector amplifiers separately.
These post-processed Stage 2 image frames were then aligned and stacked with inverse-variance weighting using the Stage 3 pipeline, keeping the original position angle of the detector for the purpose of building the PSF library.
Both F356W and F150W images were resampled with a pixel scale a factor of two smaller than that of the detector, using the drizzling algorithm implemented in the \textsf{Resample} step of the pipeline.
 The final pixel scales for F356W and F150W are $0\farcs{}0315$ and $0\farcs{}0153$, respectively.  
\\

\noindent {\bf 2D image decomposition of quasar and host galaxy emission using \galight}

\noindent Accurate decomposition of the image of a quasar into a central point source and an extended host galaxy requires a high-quality model for the PSF based on stars detected in the same image as the quasars. 
Space-based telescopes have a much sharper and more stable PSF than ground-based telescopes, and the {\it Hubble Space Telescope} (\hst) has been used to measure quasar host galaxies to redshift up to $z\sim2$ (ref. \citenum{Ding2020,  2004ApJ...614..568J, 2016ApJ...830..156M, 2019ApJ...882..141M}). 
However, \hst's $\sim$90-minute orbit means that it is continually passing between Earth's shadow and direct sunlight, causing the telescope to expand and contract (``orbital breathing'') and giving rise to a time-dependent PSF. The difficulty of modeling the PSF has not allowed quasar host galaxies to be detected significantly beyond  $z\gtrsim3$ (ref. \citenum{Glikman2015, Mechtley2012, Chen2023,Marshall2020}), where host starlight can be only measured through the SED fitting method\cite{Dong2016}. 

Here, we follow our tested strategy\cite{Ding2020}, including the analysis of JWST images\cite{2022ApJ...939L..28D}, and build a PSF library by identifying all isolated, unsaturated stars of sufficient signal-to noise ratio (SNR) in our images. 
We identified 12$|$16 PSF stars in filter F150W$|$F356W for \targettwo\ and 9$|$5 PSF stars in filter F150W$|$F356W in the images for \targetone.
We use our two-dimensional modeling software \galight\cite{Ding2020} to fit the quasar images with a model of a scaled PSF (the spatially unresolved point-like quasar) and a PSF-convolved 2D \sersic\ profile (the host galaxy). \galight\ is a \texttt{python}-based open-source package\footnote{https://galight.readthedocs.io/en/latest/} that utilizes the image modeling capabilities of \texttt{lenstronomy}\cite{2021JOSS....6.3283B}. It has been adopted for quasar decomposition of images by \hst\cite{Ding2020}, \jw\cite{2022ApJ...939L..28D}, and the HSC survey\cite{Li2021a}.
We adopt uniform priors for the effective radius \reff~$\in[0\farcs{}03 ,2\farcs{0}]$ and the \sersic\ index ($n)\in[0.3,9]$ of the host to avoid unphysical parameter inference. For \targetone, there is one nearby object which we fit simultaneously with a \sersic\ profile, allowing us to correct for its effect on the host galaxy properties.

We obtain a weighted inference for the decomposition result as briefly described here. After subtracting the remaining local background, we use each PSF in our library in turn to fit the image. The performance of each PSF is determined by its best-fit $\chi^2$ value. We select groups of two, three, and five PSFs from the library that have the top level $\chi^2$ performance and then average them using \texttt{psfr} (Birrer et al. in prep).
To optimize our modeling of the unresolved quasar emission, we consider both best-fit models using individual stars and average models based on the combined PSF stars described above. Thus, we add the three averaged PSFs as new members to the PSF library. We take the results from the five top-performing PSF models in the updated library. We determine our final result parameters by weighting their $\chi ^2$ values, as defined by ref. \citenum{Ding2020}, Eq. (3)-(6).  The uncertainties we quote are based on the dispersion in the host properties from these five different PSF models. Errors based on variation in the PSF models are often significantly larger than the formal fitting errors, as has been recently demonstrated using simulations and observations of quasars and their host galaxies\cite{2023arXiv230413776Z}. 
The final inferred host filter magnitude, size, \sersic\ index, and other fit parameters are presented in Extended Data Table~\ref{tab:table1}. The images in Figure~\ref{fig:decomp} are based on the PSF with the best performance.
The sizes of our host galaxies are defined using the \sersic\ effective radius  \reff\ along the semi-major axis measured by \galight.

The quasar host of \targettwo\ appears to be a compact galaxy. As a result, we find that the \sersic\ $n$ is poorly constrained, and the resulting host residual is point-like with an inferred \reff\ of $0\farcs{}03$ (the lower limit of the prior), indicating that the central PSF has not been fully removed. 
This model implies a stellar mass with an implausibly high value above $10^{11.8}$~$M_\odot$ given the SED fitting described below. Thus, we refit this object, fixing $n=1$ in both F356W and F150W. This gives a much-improved model. The host is quite elongated in F356W, suggesting that it is a disky edge-on galaxy. We also find a consistent position angle between F356W and F150W, as shown in Figure~\ref{fig:decomp}. Thus, for \targettwo, we adopt the results with \sersic\ $n$ fixed to 1. We further directly fit the ``data $-$ quasar" residual image (i.e., host) with the \sersic\ model and allow the \sersic\ index to vary.  When we do so, the inferred $n$ is 1.55 with values of \reff\ and the host magnitude consistent with those with the $n=1$ fit.  

Since \targetone\ has no clear host detection in the F150W band, we re-model the data and fix the host galaxy parameters (host \sersic\ index, \reff, ellipticity, and position angle) to the values inferred from the F356W band, while allowing the central position and amplitude to vary. We still see no evidence for a host residual, but the model gives a formal host magnitude of $26.3\pm0.2$ mag. We quote these inferences in Extended Data Table~\ref{tab:table1} and adopt a lower limit to the magnitude of $26.3$ mag.
\\

\noindent {\bf Confirmation of our 2D decomposition using \texttt{galfit}}

\noindent In the fitting routine described above, the results are sensitive to the setting of certain parameters. For instance, a sub-pixel shift of the PSF is typically needed to optimally align with the position of the quasar. This interpolation within a pixel can result in modest variations in the allocation of flux between the quasar and its host galaxy. In \galight, the parameter \texttt{point\_source\_supersampling\_factor} controls this interpolation. 
For guidance here on the optimal setting, we compared our results with the decomposition from \texttt{galfit}\cite{Peng2002} while implementing different supersampling factor values, i.e., 1 (no interpolation) and 2. The choice of supersampling factor had a greater impact on the F150W decomposition than that on the F356W image. We found that the two codes are in close agreement when using a \texttt{supersampling\_factor} of 2 as shown in Extended Data Figure~\ref{fig:galight_vs_galfit} and reported in Extended Data Table~\ref{tab:table1}. As a result, to account for the uncertainty associated with this setting, we report the averaged results obtained using both \texttt{supersampling\_factor} values of 1 and 2, with uncertainties covered by these two settings.
\\

\noindent \textbf{\bf Tests of the robustness of the host galaxy properties}

\noindent We perform several sanity checks for our two quasars to confirm that the detections of their host galaxies are real. First, our final reduced data are co-added using 16 dithers. To check whether 
the apparent host is caused by a random ghost from any particular dither frame, we reanalyze the data using the first 8 and the second 8 dithers separately. We are able to detect the host from both halves of the data. To ensure that the apparent host galaxy flux is not dominated by a mismatched PSF core, we mask the quasar center (using a $0\farcs{}12$ and $0\farcs{}06$ radius aperture for F356W and F150W, respectively) and redo the fit. Again, the host is well-detected. 

To rule out the possibility that the residual emission (host galaxy) is an artifact of the chosen PSF stars, we selected the five top-performing PSF in the library, and identified twenty pairs from these. We then fit one PSF with the other without assuming any host component for all pairs. We performed this test for stars in the fields of \targettwo\ and \targetone\ in both bands; residual images show no extended features in any case.  This is illustrated for \targetone\ in the F356W filter in the top panels of Extended Data Figure~\ref{fig:sanity}. We also fit galaxy + host models to pure PSF images, in which case the inferred host magnitudes were roughly two magnitudes fainter than the hosts detected around the quasars. In contrast, we clearly detect the quasar host when we use any of the five library PSFs and, using these stellar residual maps to measure the standard deviation, we display the maps of the host signal-to-noise ratio for all images of our two quasars in the bottom panels of Extended Data Figure~\ref{fig:sanity}.

We carried out three sets of simulations to further validate our host detections, including \targetone\ in the F356W filter and \targettwo\ in the F356W and F150W filters. 
In one realization of the simulation, we randomly select a PSF from the five top-ranked PSFs in the library, and add a \sersic\ profile using the best-fit parameters from our model inference.  We sample the resulting  mock quasar$+$host image at the observed resolution and insert it into a random empty position in the JWST NIRCam data. We then model the images with the other four PSF stars, and select the result with the best performance. We repeated this simulation a hundred times to quantify the bias and standard deviation of the inferred values around the true values.  We found that the key \sersic\ parameters can be obtained accurately, with biases well below 1$\sigma$ in all quantities.  
For \targetone, the bias (obtained$-$truth) and scatter of the inferred host magnitude are $-$0.02$\pm$0.11. 
For \targettwo, the magnitude bias and scatter are 0.01$\pm$0.14 and $-$0.13$\pm$0.16 for F356W and F150W, respectively. 
We present the distributions of the inferred host magnitudes in the simulations of \targetone\ (F356W) in Extended Data Figure~\ref{fig:simulation}. We note that these simulation tests were conducted with \texttt{point\_source\_supersampling\_factor} set to 1. The similar scatter distribution can be found when setting this parameter to 2.

Finally, we tested the fidelity of the host magnitude inference using a joint fit across the two bands. We fix the host galaxy parameters for \targettwo\ (where the host is detected in F150W), setting the \sersic\ parameters to those inferred from F356W; we find a very similar host magnitude to the value when all parameters are allowed to vary: the changes in the inferred host magnitude are within 0.1 mag.
\\

\noindent {\bf Stellar mass of the host galaxy} 

\noindent We use SED fitting to estimate the stellar masses of our two host galaxies at $z\sim6.4$ using the photometry and its uncertainty based on NIRCam imaging with two filters (F356W and F150W) which straddle the rest-frame 4000 \AA\ break. 
The host galaxy of \targetone\ is not detected in the F150W filter, thus we take $26.3$ mag as a lower limit on the magnitude. 
We adopt a Chabrier IMF in our analysis to allow direct comparison with 
the local black hole - host galaxy mass relations (e.g., ref. \citenum{HR2004, Bennert2011}).
Contributions of nebular emission lines from the host galaxies are also allowed with a uniform log ionization parameter $\log U$ over the range $[-3,-1]$. Three key parameters define the shape of the SED template: age, metallicity, and dust attenuation (\av). We adopt a uniform prior on the age over the range [0.01, 0.84] (in Gyr), extending to the age of the universe at $z = 6.4$. 
We adopt a uniform prior in metallicity $\log Z/Z_\odot$ over the range $[-1,-0.3]$ following  ref. \citenum{Bouwens2014, 2021arXiv211202115U}. 
We assume an \av\ range of [0, 5.0] mag for both targets. For \targetone, the F150W band has an upper limit on the flux thus the constraint on \av\ is weaker; hence, we assume \av\ follows a log-normal distribution with a median value at 0.85 as supported by recent JWST observations of high-redshift galaxies\cite{2023arXiv230103241S}. A measurement of the Balmer decrement (i.e., the ratio of narrow H$\alpha$ and H$\beta$ in the spectrum) implies an attenuation \av=0.67$\pm$0.13~mag for the narrow-line region (Onoue et al., in prep), consistent with the median extinction value of our prior.
We also note that for \targettwo, the contribution of the host in the spectrum (Onoue et al., in prep) has a magnitude in close agreement with the NIRCam photometry, supporting our stellar mass estimate.

We use \texttt{gsf}\cite{gsf} to perform the SED model fitting. This software generates a set of templates with a range of ages and metallicities according to the prior to fit a composite stellar population (CSP)-like star formation history.
A random parameter sampling is performed through Markov-Chain Monte-Carlo (MCMC) to infer the probability distribution of the SED parameters. For \targettwo\ and \targetone, the inferred values of log M$_{*}$ are $11.12\substack{+0.40\\-0.27}$ and $10.53\substack{+0.51\\-0.37}$ (in units of $M_\odot$), respectively, as shown in Extended Data Table~\ref{tab:table1}.
In Extended Data Figure~\ref{fig:sed}, we present the best-fit SEDs of the two quasar hosts, in which the MCMC inference indicates that the stellar mass is mildly sensitive to age and metallicity but is more sensitive to \av. 
We also adopt the same configuration for the independent code \texttt{Bagpipes}\cite{bagpipes} and \texttt{CIGALE}\cite{cigale} to perform SED fits; we find values of M$_{*}$ that change by 0.1 dex or less.
The robust measurement of stellar mass is due to having photometry on both sides of the 4000~\AA\ break, accurate spectroscopic redshifts, and a firm upper limit of the stellar age 0.84 Gyr, given the age of the universe at $z = 6.4$. Even so, we note that additional systematic uncertainties may be unaccounted for thus leading to greater uncertainty in the mass estimates than we have quoted.
\\

\noindent {\bf NIRSpec Observations and data reduction} 

\noindent The NIRSpec Fixed-Slit data for the two quasars were obtained on October 30, 2022 for \targettwo\ and October 28, 2022 for \targetone.
The targets were aligned onto the $0\farcs{}2\times 3\farcs{}2$ S200A2 slit and observed with the medium-resolution ($R\sim1000$) G395M grating, which covers 2.87 -- 5.27 $\mu$m in the observed frame.
Total exposures of 1,970 seconds were obtained for each target.

The spectroscopic data were processed with the JWST pipeline version 1.8.5.
Raw  (\textit{uncal})  data were downloaded from the MAST archive, which has the parameter reference files \textsf{jwst\_1027.pmap}.
The stage 1 calibration was performed with the default parameters but with the function to detect large cosmic-ray hits on the detector turned on during the \textsf{Jump} step.
The $1/f$ noise along the vertical direction was subtracted from the Stage 1 products using the same method as that for NIRCam images.
Pixels of the science frames flagged as bad or saturated were masked based on the associated Data Quality frames.
During the Stage 2 data processing, including flat-fielding and flux calibration,
we use in-flight measurements of spectrophotometric standard stars (\textsf{jwst\_nirspec\_fflat\_0065.fits}) to convert the detector signals into physical units.
The signal loss due to the optical system and the finite width of the S200A2 slit were corrected in the \textsf{Pathloss} step by applying built-in correction factors for a point source as a function of wavelength.
The two-dimensional spectra at each dither position were then stacked using the Stage 3 pipeline.
The pixel resampling scale was kept to the original pixel scale of the detector.
A 6-pixel-wide box-car aperture (0\farcs{}6 wide) was employed to extract the one-dimensional spectrum from the stacked spectrum.
\\

\noindent {\bf Spectrum fitting}

\noindent We use a public code \texttt{QSOFitMORE}\cite{Fu2021} (version 1.2.0\footnote{\url{https://doi.org/10.5281/zenodo.4893646}}) with custom modification to fit the NIRSpec rest-optical spectra of the two quasars.
The continuum emission is fitted with two components: power-law emission from the accretion disk, and a pseudo-continuum from the forest of singly-ionized iron emission.
\texttt{QSOFitMORE}  uses an empirical iron template of ref. \citenum{Boroson_Green1992}. 
The rest-frame 5100 \AA\ monochromatic luminosity ($L_{5100}$) is derived from the best-fit power-law continuum model with a power-law index of $\alpha_\lambda$ (i.e., flux density $F_\lambda\propto \lambda^{\alpha_\lambda}$).

Emission lines are modeled with a combination of one or more Gaussian profiles.
\targettwo\ needs one broad component for H$\beta$, while the adjacent forbidden emission of [O~{\sc iii}] $\lambda\lambda 4960,5008$ requires one narrow component and another blueshifted broad component.
The narrow [O~{\sc iii}] doublet is forced to have the same line width without velocity offsets between the two.
The [O~{\sc iii}] redshifts of the two targets are derived from the peaks of the narrow [O~{\sc iii}] components.
The [O~{\sc iii}] redshift of \targettwo\ agrees well with the location of the absorption lines we detect in H$\gamma$, H$\delta$, and H$\epsilon$ (Figure~\ref{fig:NIRSpecData}).
The H$\beta$+[O~{\sc iii}] line profile as well as the continuum properties of \targettwo\ are reported in Extended Data Table~\ref{tab:table2}.
A full description of the fitting procedure and other emission line properties will be presented in Onoue et al. (in prep.).
\\

\noindent {\bf BH mass estimates} 

\noindent The central black hole masses \mbh\ of the two quasars are measured by 
 the single-epoch method using broad H$\beta$ emission lines.
Assuming that the broad H$\beta$ emission traces the kinematics of the gas orbiting in the vicinity of the SMBHs (the so-called broad-line-region), one can use the line width of H$\beta$ as a proxy of the rotation velocity of the broad-line-region gas.
There is also an empirical correlation between the rest-frame 5100 \AA\ continuum luminosity ($L_{5100}$) and the radius from the SMBH to the broad-line-region gas, which is observed by the reverberation mapping technique\cite{Kaspi2000}.
In this paper, we use the recipe provided by ref. \citenum{Vestergaard2006} to estimate the black hole mass:
\begin{equation} \label{eq: MBH}
{\rm M}_{\rm BH} = 10^{6.91}  \left( \frac{L_{5100}}{10^{44}\ {\rm erg\ s^{-1}}}\right)^{0.50}  \left(\frac{{\rm FWHM_{\rm H\beta}}}{10^3\ {\rm km\ s^{-1}}} \right)^{2} M_\odot.
\end{equation}

Based on the spectrum fitting results described above, we derive virial BH masses of M$_{\rm BH} = (1.54 \pm 0.27) \times 10^9\ M_\odot$ for \targettwo\ and M$_{\rm BH} = (2.02 \pm 0.17) \times 10^8\ M_\odot$ for \targetone.
Note that these virial SMBH masses have intrinsic uncertainties of 0.4~dex\cite{Vestergaard2006}, which is not included in the error budget above.
We report the derived BH properties in Extended Data Table~\ref{tab:table2}.

We estimate the host galaxy contribution to the rest-optical continua of the NIRSpec spectra, using the decomposed host galaxy images in the NIRCam's F356W filter.  
Referring to the relative position angle of the NIRCam and NIRSpec observations,
 we find that the stellar emission falling inside the $0\farcs2 \times 0\farcs6$ extraction aperture is 21.8\% (\targettwo) and 4.8\%\ (\targetone) with respect to the total emission.
We interpret this as the fraction of the host starlight in the observed continuum emission in the NIRSpec G395M spectrum for each source, which allows us to correct the 5100 \AA\ continuum luminosity to that from the quasar emission alone.
Accordingly, the virial BH masses become M$_{\rm BH} = (1.36 \pm 0.15) \times 10^9\ M_\odot$ for \targettwo\ and M$_{\rm BH} = (1.97 \pm 0.17) \times 10^8\ M_\odot$ for \targetone, respectively.
These corrected BH properties are reported in Extended Data Table~\ref{tab:table2}, and are adopted in Figure~\ref{fig:Mbh_Mstellar}.
This correction is more significant for \targettwo, in which  stellar absorption lines are detected (Figure~\ref{fig:NIRSpecData}).
\\

\noindent {\bf Emission from H$\beta$+[O~{\sc iii}] in the hosts} 

\noindent Our SED model allows for the inclusion of the H$\beta$+[O~{\sc iii}] line emission that falls within the F365W filter (Extended Data Figure~\ref{fig:sed}).
However, if those lines were of high equivalent width in the hosts, they could affect the broad-band photometry used in the stellar mass inference, especially if the gas is ionized by the quasar\cite{Kashino2022, Matthee2022}. To ensure that our NIRCam detection of extended host emission is not dominated by these lines, we directly compare the total flux of the H$\beta$+[O~{\sc iii}] narrow lines seen in our spectra with the contribution from the host in the NIRCam image within the aperture of the spectral slit.  For \targettwo, narrow H$\beta$ emission is not detected and the [O~{\sc iii}] emission in NIRSpec contributes only $\sim 3\%$ of the host galaxy light.  

For \targetone, the narrow H$\beta$+[O~{\sc iii}] emission is comparable to the host galaxy brightness measured in the F356W filter. Much of this emission may be coming from the narrow-line region of the quasar itself. To quantify the impact on the detection of the stellar continuum, we use the 2D NIRSpec spectroscopic data and fit a Gaussian to the unresolved spatial profile at each wavelength, allowing the centroid to vary linearly with wavelength, setting the width to be the same as the spatial PSF as measured from a publicly available reference star in the same spectroscopic configuration as our observations, and letting the amplitude be a free parameter. The best-fit Gaussian profile at each wavelength is then subtracted from the 2D spectrogram to reveal any spatially extended [O~{\sc iii}] emission in the spectrum. A background region of equivalent size is subtracted from the extended emission region. We calculate the flux density of the extended emission in the F356W band and compare it to the flux density of the NIRCam host galaxy within the aperture of the NIRSpec slit on the sky. For \targetone, the [O~{\sc iii}] extended emission has a magnitude of 27.6, which is $11\%$ of the flux of the observed host galaxy within the slit aperture.
Note that the measurement of the extended emission is based on the observational fact that the apparent minor axis of \targetone\ is wider than the scale of the PSF in F356W (see Figure~\ref{fig:decomp}).
We thus conclude that the majority of the narrow H$\beta$+[O~{\sc iii}] emission in \targetone\ comes from the unresolved quasar, which has been removed by our quasar decomposition. For \targettwo, we use the same approach and find that the extended [O~{\sc iii}] has a magnitude of 27.5 mag.  The flux ratio between the extended [O~{\sc iii}] and the host galaxy within the slit is $\sim3$\%.  No extended H$\beta$ emission was detected in either target.
\\

\noindent {\bf Construction of mock quasar and host galaxy sample at $\mathbf{z=6.4}$} 

\noindent We use the method described in ref. \citenum{Li2022} to generate mock samples to assess the expected location of our two quasars in the black hole - stellar mass plane. We start with a large galaxy sample representative of the population at $z\sim6$ based on the stellar mass function\cite{Grazian2015}. We then assume that the ratio of black hole to host stellar mass and its dispersion are the same as the local values (i.e., no evolution). As an update on this procedure, we now use the BH Eddington rate distribution of ref. \citenum{Wu2022} to assign quasar luminosities. Observational uncertainty on M$_{*}$ and intrinsic uncertainty on M$_{\rm BH}$ are incorporated with a selection on magnitude ($-23.9< M_{1450} < -23.8$) is applied. The orange contours in Figure~\ref{fig:Mbh_Mstellar} show the resulting bivariate probability distribution in the stellar mass-black hole mass plane. We further generate samples for different evolution rates (M$_{\rm BH}$/M$_* \propto (1+z)^\gamma$) in steps of 0.1 in $\gamma$ and under the assumption that the dispersion is the same as the local value (ref. \citenum{Li2021b,Li2022} demonstrates that the evolution rate and dispersion are degenerate). The probability distribution of $\gamma$ is determined given our measurements of both M$_{\rm BH}$ and M$_*$ of our two quasars (Extended Data Figure~\ref{fig:gamma_prob}); $\gamma$ lies in the range $-0.40$ to $+0.84$ with 68\% confidence.

\begingroup

\renewcommand{\section}[2]{}

\endgroup

\section*{Code Availability}
The JWST data were processed with the JWST calibration pipeline (\url{https://jwst-pipeline.readthedocs.io}).
Public tools were used for data analysis: \galight\cite{Ding2020}, \texttt{galfit}\cite{Peng2002} and \texttt{PyQSOFit}\cite{Fu2021}.

\section*{Data Availability}
The JWST data are available in the MAST and can be accessed via doi:10.17909/2z59-3v32 (JWST GO 1967) after a twelve-month exclusive access period.
Reduced quasar NIRCam image cutouts are also available from the corresponding author upon request or at \url{https://github.com/dartoon/publication/tree/main/SHELLQs_J2236_J2255_data}.

\section*{Acknowledgements} 
We thank Takahiro Morishita and Lilan Yang for help with the \texttt{gsf} and \texttt{Bagpipes} packages.
We thank Yuming Fu for his help on the use of \texttt{QSOFitMORE}.
We thank Xiaohui Fan, Luis Ho, and Vardha Bennert for useful discussions.

This work is based on observations made with the NASA/ESA/CSA James Webb Space Telescope. The data were obtained from the Mikulski Archive for Space Telescopes at the Space Telescope Science Institute, which is operated by the Association of Universities for Research in Astronomy, Inc., under NASA contract NAS 5-03127 for JWST. These observations are associated with program \#1967. Support for program \#1967 was provided by NASA through a grant from the Space Telescope Science Institute, which is operated by the Association of Universities for Research in Astronomy, Inc., under NASA contract NAS 5-03127.

This paper is based on data collected at the Subaru Telescope and retrieved from the HSC data archive system, which is operated by the Subaru Telescope and Astronomy Data Center (ADC) at NAOJ. Data analysis was in part carried out with the cooperation of Center for Computational Astrophysics (CfCA), NAOJ. We are honored and grateful for the opportunity of observing the Universe from Maunakea, which has the cultural, historical and natural significance in Hawaii. 

This work was supported by World Premier International Research Center Initiative (WPI), MEXT, Japan. 
XD, JS, YM, TI, KK, and HU are supported by the Japan Society for the Promotion of Science (JSPS) KAKENHI grant Nos. JP22K14071, JP18H01251, JP22H01262, JP17H04830, JP21H04494,  JP20K14531, JP17H06130, JP20H01953.
FW, SEIB, BT, JTS, and MO acknowledge support from the ERC grant Nos. 740246, 885301, 950533.
MO and KI acknowledge support from the National Natural Science Foundation of China grant Nos. 12150410307 and 12073003. 
YM acknowledges support from the Mitsubishi Foundation grant No. 30140. 
SF acknowledges support from  NASA through the NASA Hubble Fellowship grant HST-HF2-51505.001-A awarded by the STScI, which is operated by the Association of Universities for Research in Astronomy, Incorporated, under NASA contract NAS5-26555.
ZH acknowledges support from NSF grant AST-2006176.
KI acknowledges support by the Spanish MCIN under grant PID2019-105510GB-C33/AEI/10.13039/501100011033
AL acknowledges funding from MIUR under the grant PRIN 2017-MB8AEZ.
BT acknowledges support from the Israel Science Foundation (grant No. 1849/19) and from the European Research Council (ERC) under the European Union’s Horizon 2020 research and innovation program (grant agreements No. 950533).
MT acknowledges support from the NWO grant 0.16.VIDI.189.162 (``ODIN'').
MV acknowledges support from the Independent Research Fund Denmark via grant number DFF 8021-00130.

\section*{Author Contributions}
XD led the analysis of the quasar decomposition, simulation test, SED fitting, and preparation of the manuscript. 
MO reduced the NIRCam and NIRSpec data. He also led the spectroscopic data analysis and wrote relevant sections.
We regard that these first two authors equally contributed to this work.
JS contributed to the NIRCam data analysis and preparation of the manuscript. CP performed the NIRSpec 2D spectroscopic analysis and evaluated the strength of the extended [O~{\sc iii}] emission. JL constructed the mock quasar sample at $z=6.4$, which was used to assess the redshift evolution of the black hole -- stellar mass relation.
MO, XD, JS, YM, TI, MS, and KJ led the project design and management, also developing the main interpretation of the results.  
All co-authors contributed to the discussion of the presented results and the preparation of the manuscript.

\section*{Competing interests} The authors declare no competing interests.

\section*{Correspondence} Correspondence and requests for materials should be addressed to X. Ding (\url{xuheng.ding@ipmu.jp}) and M. Onoue (\url{onoue@pku.edu.cn}).

\clearpage

\section*{Extended Data}

\captionsetup[table]{name=Extended Data Table}

\begin{table}[ht]
\begin{center}
\begin{tabular}{lcccc}
\hline \noalign {\smallskip}
 & \multicolumn{2}{c}{\targettwo} & \multicolumn{2}{c}{\targetone} \\
\hline 
RA & \multicolumn{2}{c}{22 36 44.58} & \multicolumn{2}{c}{22 55 38.04} \\
Dec & \multicolumn{2}{c}{+00 32 56.90}  & \multicolumn{2}{c}{+02 51 26.60}  \\
Redshift  & \multicolumn{2}{c}{6.40}  & \multicolumn{2}{c}{6.34}  \\
$M_{1450}$ (mag) & \multicolumn{2}{c}{ $-$23.8} &  \multicolumn{2}{c}{ $-$23.9 }   \\ 
\hline 
 & F356W  & F150W   & F356W  & F150W   \\
\hline 
\reff~($\arcsec$)  &  0.13$\pm$0.02 &  0.10$\pm$0.03  & 0.27$\pm$0.19 & - \\
\reff~(kpc) & 0.7$\pm$0.1 & 0.5$\pm$0.2  & 1.5$\pm$1.1 & - \\
\sersic\ index & fixed as 1 & fixed as 1 & 1.5$\pm$1.4 & - \\
ellipticity (b/a)  & 0.39$\pm$0.02 & 0.34$\pm$0.11 & 0.61$\pm$0.09 & -\\
PA (N to W) & $-$39.4$^{\circ}$$\pm$2.8$^{\circ}$ & $-$48.8$^{\circ}$$\pm$25.5$^{\circ}$  & 26.0$^{\circ}$$\pm$11.6$^{\circ}$ & - \\
flux ratio  &  25.5\%$\pm$4.4\% & 10.2\%$\pm$2.8\%  & 9.8\%$\pm$2.6\% & $<$ 3.8$\%$ \\
quasar mag & 21.95$\pm$0.07 & 22.73$\pm$0.04  & 22.15$\pm$0.03 & 22.89$\pm$0.02 \\
host mag & 23.12$\pm$0.20 & 25.12$\pm$0.29 &  24.58$\pm$0.30 & $>$ 26.3 \\
host mag (\texttt{galfit}) & 23.10$\pm$0.07 & 24.95$\pm$0.10 & 24.27$\pm$0.06 & $>$ 26.7\\
\hline 
host $\log$ M$_{*}$  ($M_{\odot}$) & \multicolumn{2}{c}{$11.12\substack{+0.40\\-0.27}$}  & \multicolumn{2}{c}{$10.53\substack{+0.51\\-0.37}$} \\
\hline 
\end{tabular}
{\small \caption{
\textbf{Summary of the quasar decomposition results.} The photometry measurements are obtained using \galight, where the top-performance PSFs are weighted with different \texttt{point\_source\_supersampling\_factor}.  
We also present the host magnitudes inferred by \texttt{galfit}, combining results using top-performance PSFs.
The RA and Dec are from the HSC measurements, consistent with \jw's. 
The stellar masses are from \texttt{gsf}; we also find consistent results using the \texttt{Bagpipes} and \texttt{CIGALE} codes.
\label{tab:table1}}}
\end{center}
 \vspace{-0.5cm}
\end{table}

\clearpage

\begin{table}[ht]
    \centering
    \begin{tabular}{lllll}
    \hline
          & J2236+0032 & (corrected) & J2255+0251 & (corrected) \\ \hline\hline
        [O~{\sc iii}] redshift & $6.4039 \pm 0.0009$ & & $6.333 \pm 0.001$ & \\ 
        $F_{\rm H\beta,broad}$  & $2.13\pm0.10$ & & $7.01\pm0.09$  & \\ 
        $F_{\rm H\beta,narrow}$  &  & & $0.36\pm0.07$ & \\ 
        $F_{\rm [O~{\sc III}] \lambda 4960,core}$  & $0.07\pm0.10$ & & $0.26\pm0.04$ & \\ 
        $F_{\rm [O~{\sc III}] \lambda 4960,wing}$ & $0.49\pm0.14$ & & $0.31\pm0.02$& \\ 
        $F_{\rm [O~{\sc III}] \lambda 5008,core}$  & $0.34 \pm 0.13$ & & $0.66 \pm 0.09$ & \\ 
        $F_{\rm [O~{\sc III}] \lambda 5008,wing}$  & $0.96 \pm 0.18$ & & $1.36 \pm 0.08$ & \\ 
        $\alpha_\lambda$ & $-1.752\pm0.003$ & & $-1.75\pm0.02$ & \\ 
        $L_{5100}$ ($10^{45}$ erg s$^{-1}$) & $2.29 \pm 0.02$ &  $1.79 \pm 0.01$  & $1.531 \pm 0.008$ &  $1.458 \pm 0.008$ \\ 
        H$\beta$ FWHM$_{\rm broad}$ (km s$^{-1}$) & $6290 \pm 560$ & & $2520 \pm 110$&  \\ 
        M$_{\rm BH}$ ($10^8\ M_\odot$) & $15.4 \pm 2.7$ & $13.6 \pm 1.5$ & $2.02 \pm 0.17$ & $1.97 \pm 0.17$   \\ 
        \hline
    \end{tabular} 
{\small \caption{
\textbf{Spectroscopic properties and virial BH masses.} The corrected values are based on the quasar continuum emission after subtracting the host galaxy light falling onto the NIRSpec fixed slit. Flux values are in units of $10^{-17}$ erg cm$^{-2}$ s$^{-1}$.
\label{tab:table2}}}
\end{table}

\captionsetup[figure]{name=Extended Data Figure}
\setcounter{figure}{0}

\begin{figure*}[ht]
\centering
\includegraphics[width=0.8\textwidth]{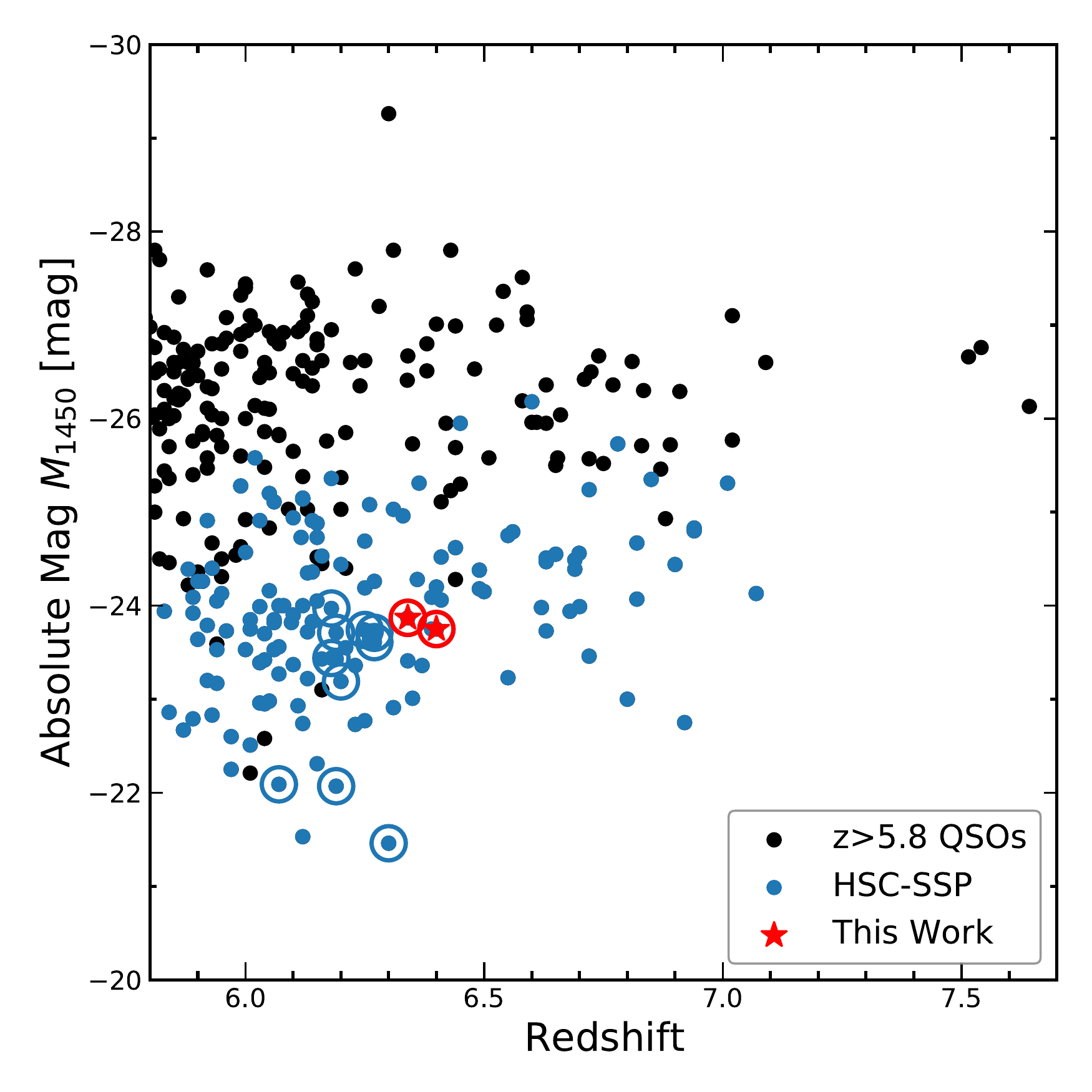}\\
\caption{\textbf{Distribution of known quasars at redshift $\mathbf{z>5.8}$.}
The y-axis indicates the absolute magnitudes at rest-frame 1450 \AA.
The two target quasars in this work are shown in red (\targettwo\ at $z=6.40$, and \targetone\ at $z=6.34$), while other low-luminosity quasars from the HSC-SSP are shown in blue.
The \jw\ 12 Cycle 1 targets in GO \#1967 are highlighted with open circles.
Other known quasars are shown in black. \label{fig:shellqs_sample}}
\end{figure*}

\begin{figure*}[ht]
\centering
\includegraphics[trim = 0mm 0mm 0mm 0mm, clip, width=0.85\textwidth]{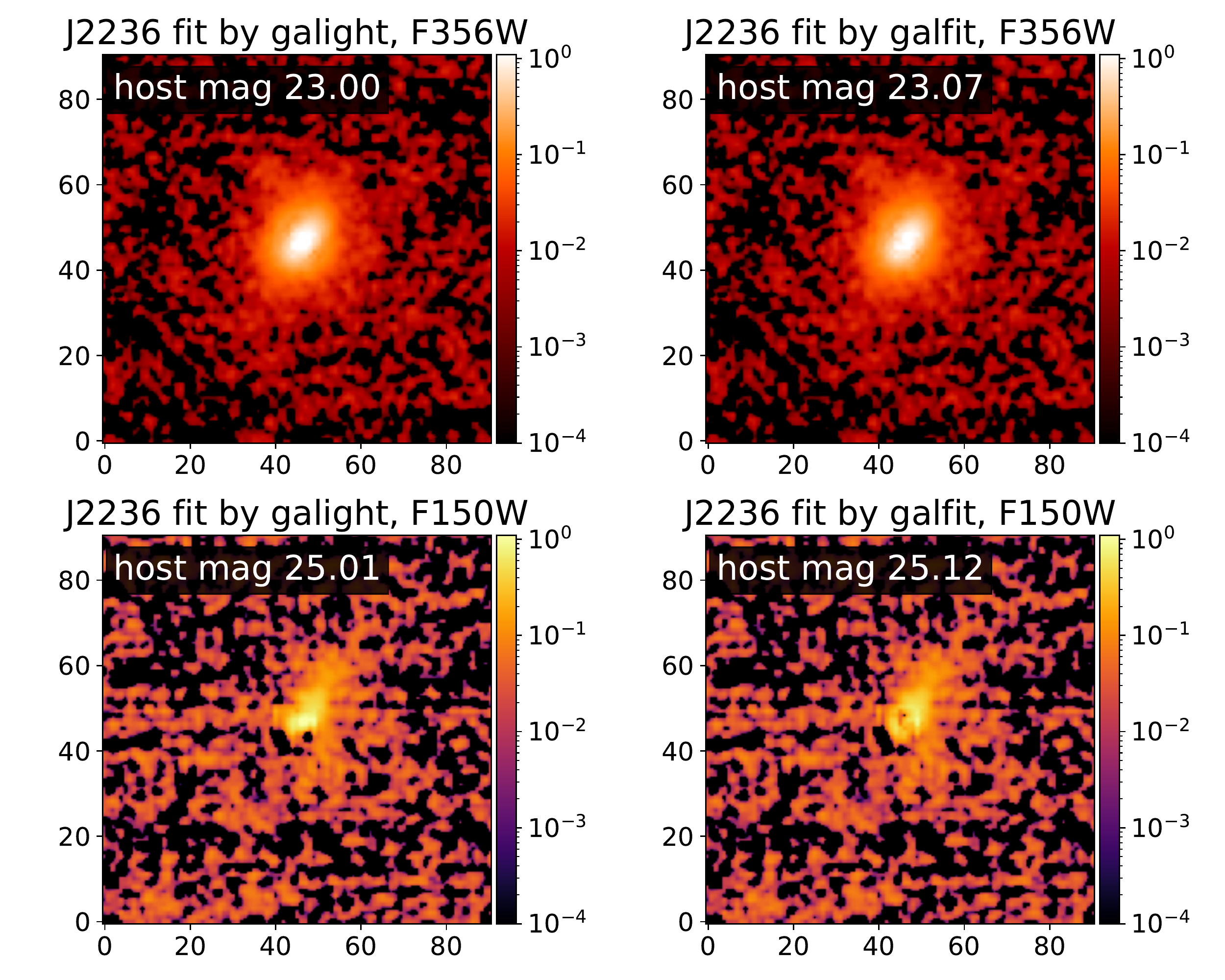}\\
\caption{\textbf{Quasar-subtracted images of \targettwo\ 
based on 2D decomposition by two independent codes}. The inferred host magnitude from each code is indicated. For this comparison, the \texttt{point\_source\_supersampling\_factor} is set to 2 for \galight. The same PSF is used for this comparison.
Coordinates are in pixel units; the pixel scales  for F356W and F150W are $0\farcs{}0315$ and $0\farcs{}0153$, respectively.
\label{fig:galight_vs_galfit}}
\end{figure*}

\begin{figure*}[ht]
\centering
\includegraphics[trim = 0mm 0mm 0mm 0mm, clip, width=0.9\textwidth]{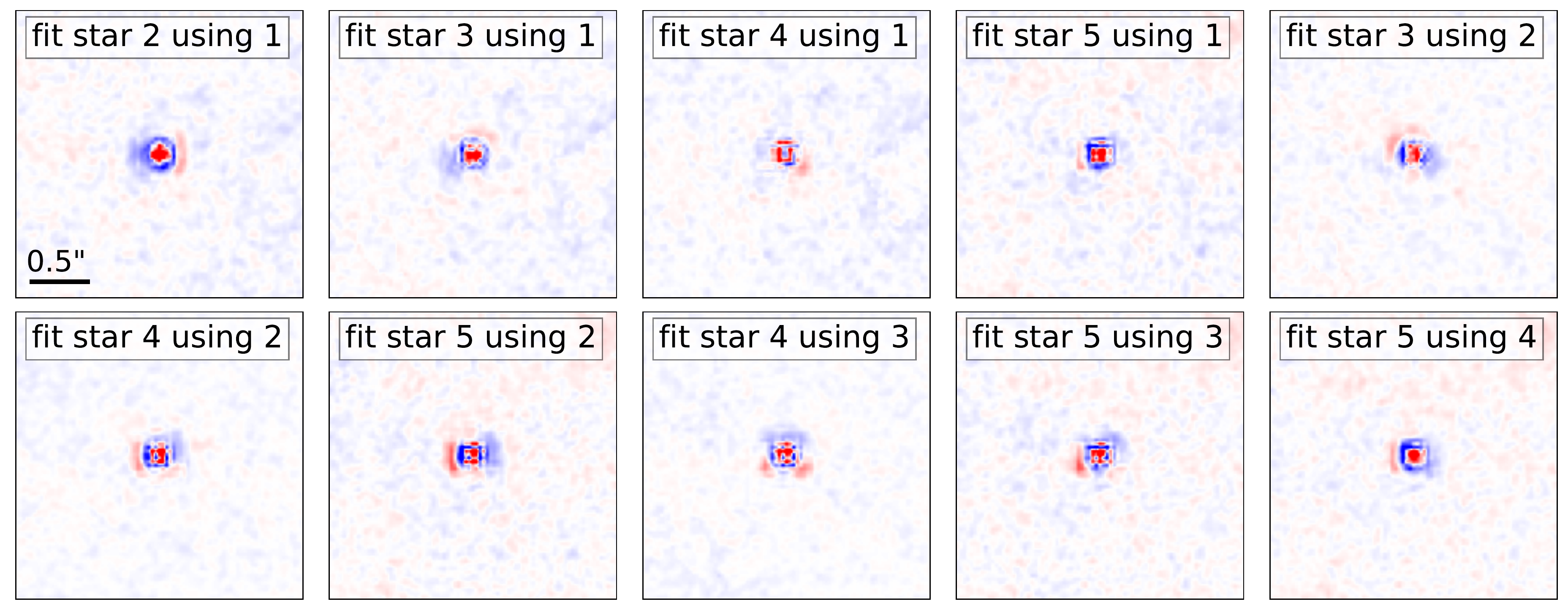}\\
\includegraphics[trim = 0mm 78mm 0mm 40mm, clip, width=0.9\textwidth]{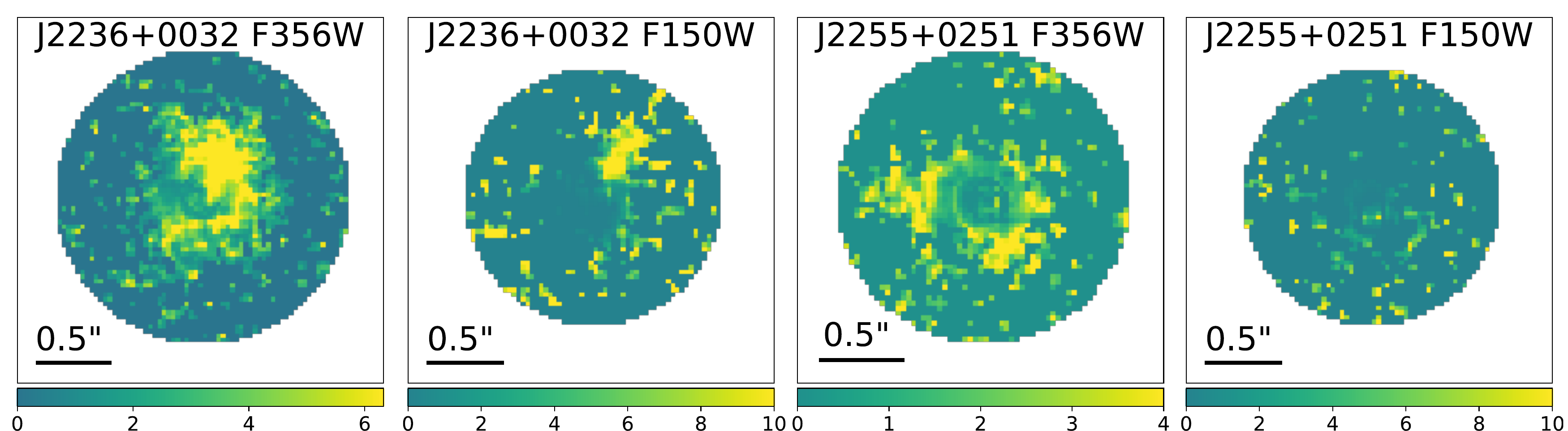}

\caption{
\textbf{Testing the effect of PSF star mismatch on the inferred presence of a host galaxy.} {\it Top:} The residuals using the top five PSF star pairs for \targetone\ in F356W (10 of 20 are shown) -- we see no evidence for extended residual emission. Using these 20 PSF residual maps, the PSF mismatch {\em standard deviation} is calculated for each pixel. {\it Bottom:} Maps of host galaxy signal (i.e., data$-$point-source model) divided by the standard deviation (as indicated by the color scale) for both targets in two filters.  The host is clearly detected in all cases except for \targetone\ in F150W. \label{fig:sanity}}
\end{figure*}

\begin{figure*}[ht]
\centering
\includegraphics[trim = 0mm 0mm 0mm 0mm, clip, width=0.95\textwidth]{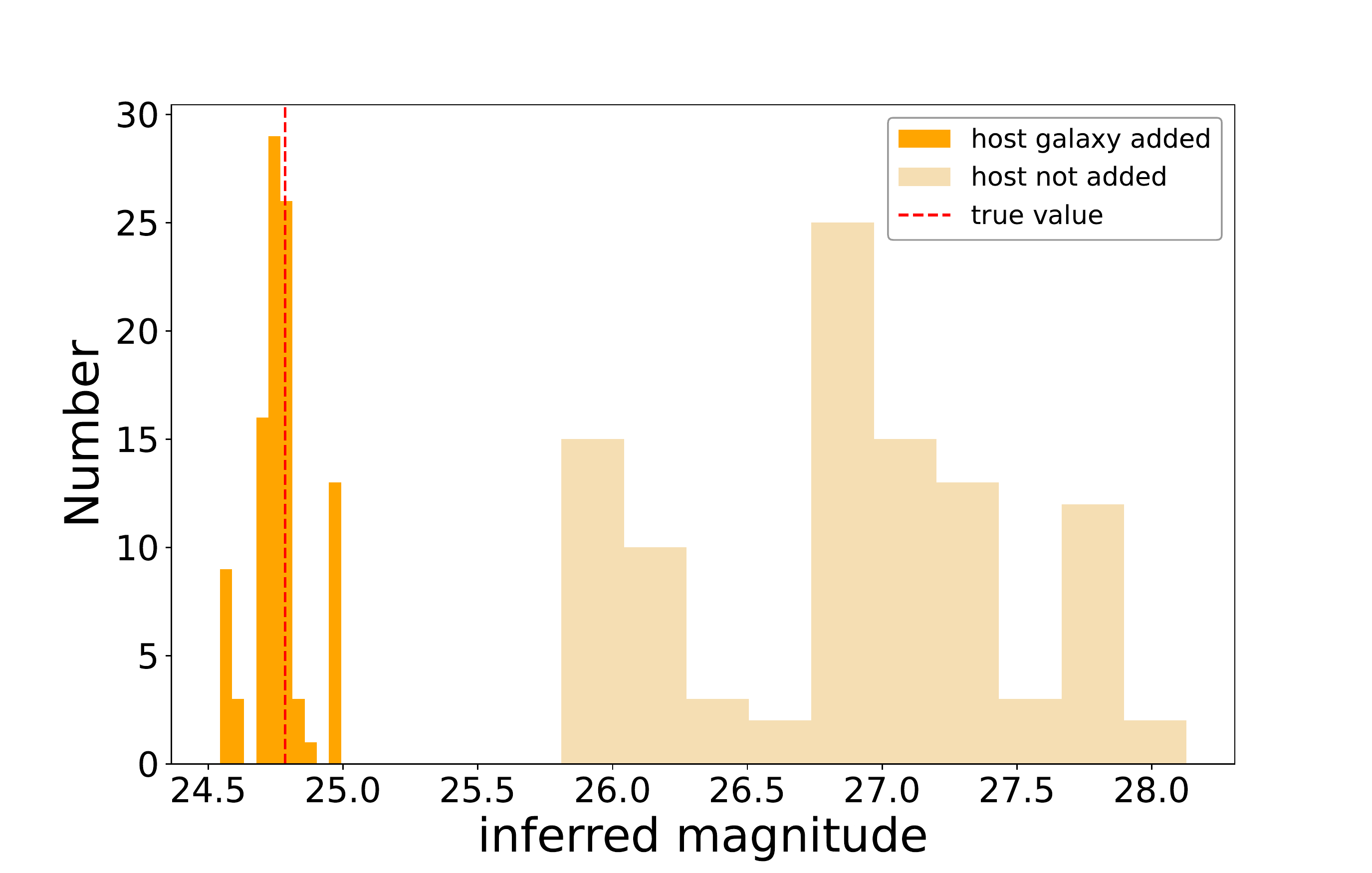}\\
\caption{\textbf{Distribution of the inferred host magnitude from simulations with and without a host present.}
When the host of \targetone\ is added into the mock image in the simulations, the derived host magnitude distribution based on 100 realizations  (dark orange) is consistent with the input true value, with an r.m.s. scatter of 0.12 mag. When the host light is not included in the mock image and we still fit the host light in our model, we infer host magnitudes with a broad distribution (tan histogram), centered two magnitudes fainter.\label{fig:simulation}}
\end{figure*}

\begin{figure*}[ht]
\centering
\includegraphics[trim = 0mm 50mm 0mm 10mm, clip, width=1\textwidth]{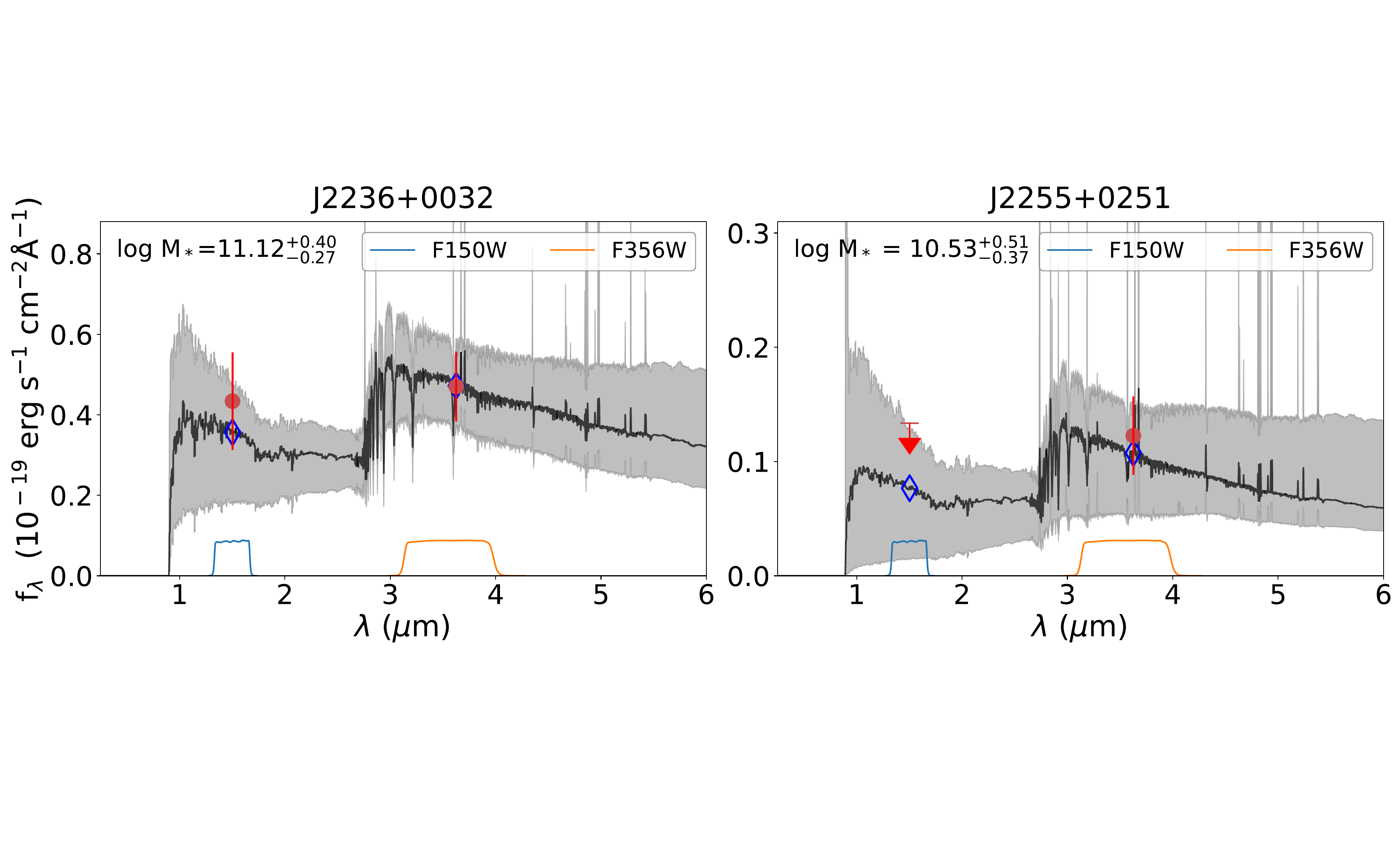}\\
\includegraphics[trim = 0mm 0mm 0mm 0mm, clip, width=1.0\textwidth]{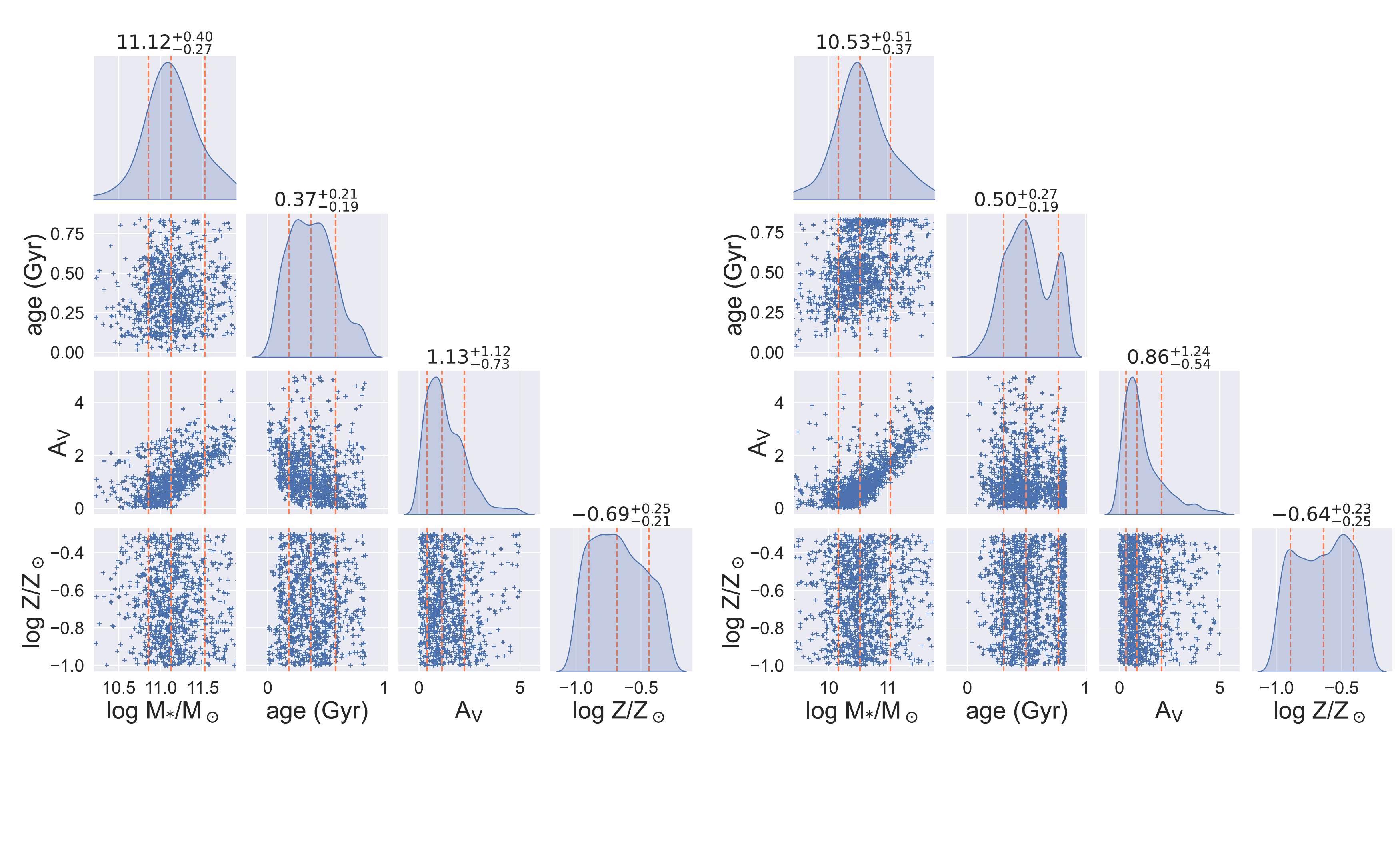}\\
\vspace{-1.5cm}
\caption{\textbf{The SED inference using the host galaxy two-band photometry based on \texttt{gsf}.} {\it Top:} The red data points with errors indicate the inferred host fluxes. The gray region represents the 1$\sigma$ range of variation in SED templates obtained through MCMC, while the black line represents the median SED template. The blue diamonds indicate the predictions using this median template. 
{\it Bottom:} The distribution of the host galaxy SED properties using MCMC inference. 
}
\label{fig:sed}
\end{figure*}

\begin{figure*}[ht]
\centering
\includegraphics[trim = 0mm 0mm 0mm 0mm, clip, width=0.75\textwidth]{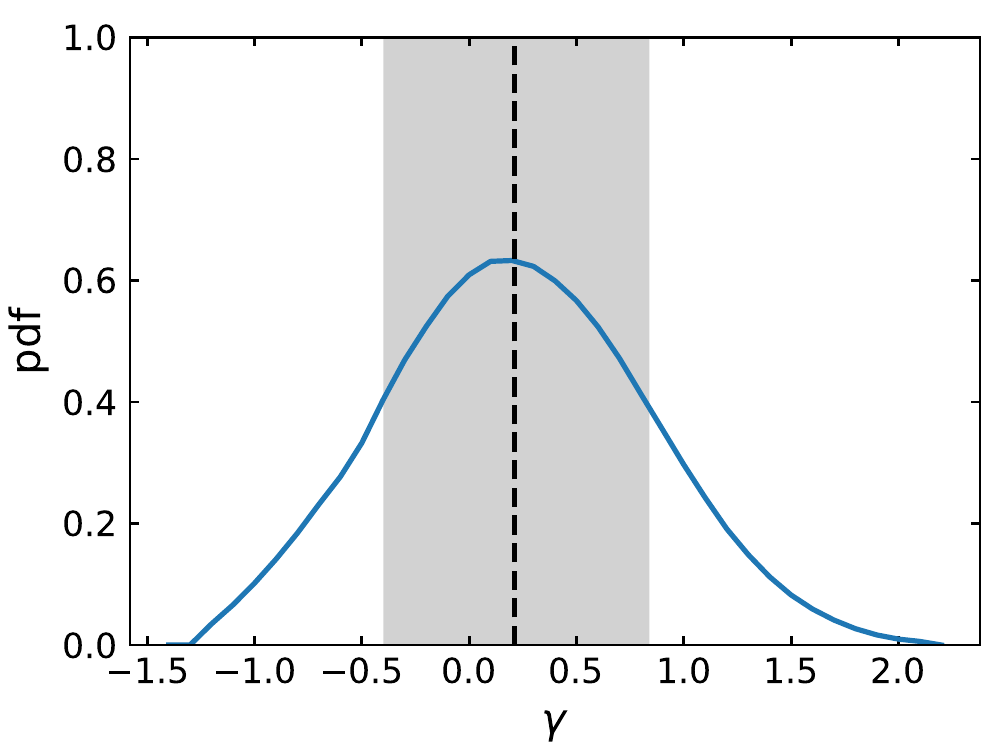}\\
\caption{\textbf{Probability distribution of the evolution parameter based on observations of our two quasars.} The evolution parameter 
 $\gamma$ is inferred as M$_{\rm BH}$/M$_* \propto (1+z)^\gamma$.
The 68\% confidence interval in $\gamma$ lies between $-0.40$ and $+0.84$.
The black dashed line indicates the 50 percentile position ($+0.21$). 
\label{fig:gamma_prob}}
\end{figure*}


\begin{thebibliography}{99}

\bibitem{Mechtley2012} Mechtley, M. et al. Near-infrared Imaging of a z = 6.42 Quasar Host Galaxy with the Hubble Space Telescope Wide Field Camera 3. \textit{\apj}\textbf{756}, L38 (2012)
\bibitem{Marshall2020} Marshall, M.~A. et al. Limits to Rest-frame Ultraviolet Emission from Far-infrared-luminous z $\sim$ 6 Quasar Hosts. \textit{\apj} \textbf{900}, 21 (2020)
\bibitem{Peng2006} Peng, C.~Y. et al. Probing the coevolution of supermassive black holes and galaxies using gravitationally lensed hosts. \textit{\apj} \textbf{649}, 616-634 (2006)
\bibitem{Matsuoka2016} Matsuoka, Y. et al. Subaru High-z Exploration of Low-luminosity Quasars (SHELLQs). I. Discovery of 15 Quasars and Bright Galaxies at $5.7 < z < 6.9$. \textit{\apj} \textbf{828}, 26 (2016)
\bibitem{Matsuoka2018} Matsuoka, Y. et al. Subaru High-z Exploration of Low-luminosity Quasars (SHELLQs). IV. Discovery of 41 Quasars and Luminous Galaxies at 5.7 {\ensuremath{\leq}} z {\ensuremath{\leq}} 6.9. \textit{\apjs} \textbf{237}, 5 (2018)
\bibitem{Matsuoka2022} Matsuoka, Y., et al. Subaru High-z Exploration of Low-luminosity Quasars (SHELLQs). XVI. 69 New Quasars at 5.8 {\ensuremath{\leq}} z {\ensuremath{\leq}} 7.0. \textit{\apjs} \textbf{259}, 18 (2022)
\bibitem{HSC-SSP2018} Aihara, H., et al. The Hyper Suprime-Cam SSP Survey: Overview and survey design. \textit{\pasj} \textbf{70}, S4 (2018)
\bibitem{Fan2001} Fan, X. et al.. A Survey of $z>5.8$ Quasars in the Sloan Digital Sky Survey. I. Discovery of Three New Quasars and the Spatial Density of Luminous Quasars at z\raisebox{-0.5ex}\textasciitilde6. \textit{\apj} \textbf{122}, 2833–2849 (2001)
\bibitem{2018Natur.553..473B} Ba{\~n}ados, E. et al. An 800-million-solar-mass black hole in a significantly neutral Universe at a redshift of 7.5. \textit{\nat} \textbf{553}, 473–476 (2018)
\bibitem{Fan2022} Fan, X., Banados, E., \& Simcoe, R.~A. Quasars and the Intergalactic Medium at Cosmic Dawn. \textit{arXiv e-prints} \textbf arXiv:2212.06907 (2022)
\bibitem{Rieke2005} Rieke, M.~J., Kelly, D. \& Horner, S.\ Overview of James Webb Space Telescope and NIRCam's Role.\ Cryogenic Optical Systems and Instruments XI. \textbf{5904}, 1–8 (2005)
\bibitem{Ding2020} Ding, X. et al.\ The Mass Relations between Supermassive Black Holes and Their Host Galaxies at $1 < z < 2$ HST-WFC3. \textit{\apj} \textbf{888}, 37 (2020)
\bibitem{Sersic1963} S{\'e}rsic, J.~L.\ Boletin de la Asociacion Argentina de Astronomia La Plata Argentina, \textbf{6}, 41-43 (1963)
\bibitem{NIRSpec} Jakobsen, P., et al. The Near-Infrared Spectrograph (NIRSpec) on the James Webb Space Telescope. I. Overview of the instrument and its capabilities. \textit{\aap} \textbf{661}, A80 (2022)
\bibitem{Grazian2015} Grazian, A., et al. The galaxy stellar mass function at 3.5 {\ensuremath{\leq}}z {\ensuremath{\leq}} 7.5 in the CANDELS/UDS, GOODS-South, and HUDF fields. \textit{\aap} \textbf{575}, A96 (2015)
\bibitem{Vestergaard2006} Vestergaard, M., \& Peterson, B.~M. Determining Central Black Hole Masses in Distant Active Galaxies and Quasars. II. Improved Optical and UV Scaling Relationships. \textit{\apj} \textbf{641}, 689-709 (2006)
\bibitem{Li2022} Li, J., et al. On the Connection between Supermassive Black Holes and Galaxy Growth in the Reionization Epoch. \textit{\apjl} \textbf{931}, L11 (2022)
\bibitem{Habouzit2022} Habouzit, M., et al. Co-evolution of massive black holes and their host galaxies at high redshift: discrepancies from six cosmological simulations and the key role of JWST. \textit{\mnras} \textbf{511}, 3751-3767 (2022)
\bibitem{2013ARA&A..51..511K} Kormendy, J., \& Ho, L.~C. Coevolution (Or Not) of Supermassive Black Holes and Host Galaxies. \textit{\araa} \textbf{51}, 511-653 (2013)
\bibitem{Inayoshi2022} Inayoshi, K., et al. Rapid Growth of Seed Black Holes during Early Bulge Formation. \textit{\apj} \textbf{927}, 237 (2022)
\bibitem{HR2004} H{\"a}ring, N., \& Rix, H.-W. On the Black Hole Mass-Bulge Mass Relation. \textit{\apjl} \textbf{604}, L89-L92 (2004)
\bibitem{Bennert2011} Bennert, V.~N., Auger, M.~W., Treu, T., Woo, J.-H., \& Malkan, M.~A. A Local Baseline of the Black Hole Mass Scaling Relations for Active Galaxies. I. Methodology and Results of Pilot Study. \textit{\apj} \textbf{726}, 59 (2011)

\end{thebibliography}

\begin{thebibliography}{99}
\bibitem[23]{Matsuoka2018c} Matsuoka, Y., et al. Subaru High-z  Exploration of Low-luminosity Quasars (SHELLQs). V. Quasar Luminosity Function and Contribution to Cosmic Reionization at z = 6. \textit{\apj} \textbf{869}, 150 (2018)
\bibitem[24]{Bradley2022} Bradley, L., Sip{\H{o}}cz, B., Robitaille, T., et al. astropy/photutils: 1.4.0 Zenodo~6385735B (2022)
\bibitem[25]{2004ApJ...614..568J} Jahnke, K., et al. Ultraviolet Light from Young Stars in GEMS Quasar Host Galaxies at 1.8$<$z$<$2.75. \textit{\apj} \textbf{614}, 568-585 (2004)
\bibitem[26]{2016ApJ...830..156M} Mechtley, M., et al. Do the Most Massive Black Holes at z = 2. Grow via Major Mergers? \textit{\apj} \textbf{830}, 156 (2016)
\bibitem[27]{2019ApJ...882..141M} Marian, V., et al. Major Mergers Are Not the Dominant Trigger for High-accretion AGNs at z {\ensuremath{\sim}} 2. \textit{\apj} \textbf{882}, 141 (2019)
\bibitem[28]{Glikman2015} Glikman, E., et al. Major Mergers Host the Most-luminous Red Quasars at z {\ensuremath{\sim}} 2: A Hubble Space Telescope WFC3/IR Study. \textit{\apj} \textbf{806}, 218 (2015)
\bibitem[29]{Chen2023} Chen, Y.-C., et al. A close quasar pair in a disk-disk galaxy merger at z = 2.17. \textit{\nat} \textbf{616}, 45-49 (2023)
\bibitem[30]{Dong2016} Dong, X.~Y., \& Wu, X.-B. Herschel Observed Stripe 82 Quasars and Their Host Galaxies: Connections between AGN Activity and host Galaxy Star Formation. \textit{\apj} \textbf{824}, 70 (2016)
\bibitem[31]{2021JOSS....6.3283B} Birrer, S., et al. lenstronomy II: A gravitational lensing software ecosystem. \textit{The Journal of Open Source Software} \textbf{6}, 3283 (2021)
\bibitem[32]{2022ApJ...939L..28D} Ding, X., Silverman, J.~D., \& Onoue, M. Opening the Era of Quasar-host Studies at High Redshift with JWST. \textit{\apjl} \textbf{939}, L28 (2022)
\bibitem[33]{Li2021a} Li, J., et al. The Sizes of Quasar Host Galaxies in the Hyper Suprime-Cam Subaru Strategic Program. \textit{\apj} \textbf{918}, 22 (2021)
\bibitem[34]{2023arXiv230413776Z} Zhuang, M.-Y., \& Shen, Y. Characterization of JWST NIRCam PSFs and Implications for AGN+Host Image Decomposition. \textit{arXiv e-prints} \textbf arXiv:2304.13776 (2023)
\bibitem[35]{Peng2002} Peng, C.~Y., Ho, L.~C., Impey, C.~D., \& Rix, H.-W. Detailed Structural Decomposition of Galaxy Images. \textit{\aj} \textbf{124}, 266-293 (2002)
\bibitem[36]{Bouwens2014} Bouwens, R.~J., et al. UV-continuum Slopes of $>$4000 z {\ensuremath{\sim}} 4-8 Galaxies from the HUDF/XDF, HUDF09, ERS, CANDELS-South, and CANDELS-North Fields. \textit{\apj} \textbf{793}, 115 (2014)
\bibitem[37]{2021arXiv211202115U} Ucci, G., et al. Astraeus - II. Quantifying the impact of cosmic variance during the Epoch of Reionization. \textit{\mnras} \textbf{506}, 202-214 (2021)
\bibitem[38]{2023arXiv230103241S} Shapley, A.~E., Sanders, R.~L., Reddy, N.~A., Topping, M.~W., \& Brammer, G.~B. JWST/NIRSpec Balmer-line Measurements of Star Formation and Dust Attenuation at z {\ensuremath{\sim}} 3-6. \textit{arXiv e-prints} \textbf arXiv:2301.03241 (2023)
\bibitem[39]{gsf} Morishita, T., et al. Massive Dead Galaxies at z {\ensuremath{\sim}} 2 with HST Grism Spectroscopy. I. Star Formation Histories and Metallicity Enrichment. \textit{\apj} \textbf{877}, 141 (2019)
\bibitem[40]{bagpipes} Carnall, A.~C., McLure, R.~J., Dunlop, J.~S., \& Dav{\'e}, R. Inferring the star formation histories of massive quiescent galaxies with BAGPIPES: evidence for multiple quenching mechanisms. \textit{\mnras} \textbf{480}, 4379-4401 (2018)
\bibitem[41]{cigale} Boquien, M., et al. CIGALE: a python Code Investigating GALaxy Emission. \textit{\aap} \textbf{622}, A103 (2019)
\bibitem[42]{Fu2021} Fu, Y. QSOFITMORE (version 1.2.0): a python package for fitting UV-optical spectra of quasars. \textit{Zenodo.5810042} \textbf (2021)
\bibitem[43]{Boroson_Green1992} Boroson, T.~A., \& Green, R.~F. The Emission-Line Properties of Low-Redshift Quasi-stellar Objects. \textit{\apjs} \textbf{80}, 109 (1992)
\bibitem[44]{Kaspi2000} Kaspi, S., et al. Reverberation Measurements for 17 Quasars and the Size-Mass-Luminosity Relations in Active Galactic Nuclei. \textit{\apj} \textbf{533}, 631-649 (2000)
\bibitem[45]{Kashino2022} Kashino, D., et al. EIGER I. a large sample of [OIII]-emitting galaxies at $5.3 < z < 6.9$ and direct evidence for local reionization by galaxies. \textit{arXiv e-prints} \textbf arXiv:2211.08254 (2022)
\bibitem[46]{Matthee2022} Matthee, J., et al. EIGER II. first spectroscopic characterisation of the young stars and ionised gas associated with strong H$\beta$ and [OIII] line-emission in galaxies at z=5-7 with JWST. \textit{arXiv e-prints} \textbf arXiv:2211.08255 (2022)
\bibitem[47]{Wu2022} Wu, J., et al. Demographics of $z \sim 6$ quasars in the black hole mass-luminosity plane. \textit{\mnras} \textbf{517}, 2659-2676 (2022)
\bibitem[48]{Li2021b} Li, J., et al. Synchronized Coevolution between Supermassive Black Holes and Galaxies over the Last Seven Billion Years as Revealed by Hyper Suprime-Cam. \textit{\apj} \textbf{922}, 142 (2021)

\end{thebibliography}
\end{document}